\documentclass[10pt]{article}
\usepackage{geometry}[scale=0.88]
\usepackage{amsmath,amssymb,graphicx,alltt}
\usepackage{float}
\usepackage{listings}
\usepackage{enumitem}
\usepackage{booktabs}
\usepackage{xcolor}
\usepackage{abstract}
\usepackage[colorlinks, linkcolor=red, anchorcolor=blue, citecolor=green]{hyperref}

\newcommand{\revised}[1]{#1}
\newcommand{\hdsdp}{\texttt{HDSDP}}
{}

\begin{document}

\title{HDSDP: Software for Semidefinite Programming}

\author{Wenzhi Gao\thanks{gwz@stanford.edu} \quad Dongdong Ge\thanks{ge.dongdong@mail.shufe.edu.cn} \quad Yinyu Ye\thanks{yyye@stanford.edu}}\maketitle

\begin{abstract}
  {{\texttt{HDSDP}}} is a numerical software solving the semidefinite programming problems. The main framework of {{\texttt{HDSDP}}} resembles the dual-scaling
  interior point solver {{\texttt{DSDP}}}{\cite{benson2008algorithm}} and several new
  features, including a dual method based on the simplified homogeneous self-dual embedding, have been implemented. The embedding technique enhances stability of the dual method 
  and several new heuristics and computational techniques are designed to accelerate its convergence. 
  HDSDP aims to show how dual-scaling algorithm benefits from the
  self-dual embedding and it is developed in parallel to {{{\texttt{DSDP5.8}}}}.
  Numerical experiments over several classical benchmark datasets exhibit its robustness and efficiency, and particularly its advantages on SDP instances featuring low-rank structure and sparsity. \revised{{\hdsdp} is open-sourced under MIT license and available at \href{https://github.com/COPT-Public/HDSDP}{https://github.com/COPT-Public/HDSDP}}.
 
\end{abstract}
\section{Introduction}

Semidefinite programming (SDP) is defined
by
\begin{eqnarray}
  \min_{\mathbf{X}} & \left\langle \mathbf{C}, \mathbf{X} \right\rangle & \nonumber \\
  \text{subject to} & \mathcal{A} \mathbf{X} = \mathbf{b} & \\ 
  & \mathbf{X} \in \mathbb{S}_+^n, & \nonumber
\end{eqnarray}
where we study linear optimization subject to affine constraints over the cone of
positive-semidefinite matrices. Due to its extensive modeling capability, SDP has been broadly adopted by communities including combinatorial optimization {\cite{goemans1995improved, laurent2005semidefinite}},
dynamic systems {\cite{vandenberghe1996semidefinite}}, sums of squares
optimization {\cite{laurent2009sums}}, quantum
information {\cite{hayashi2016quantum}}, and distance geometry
{\cite{biswas2004semidefinite, so2007theory}}. We refer the interested
readers to {\cite{wolkowicz2005semidefinite}} for a more comprehensive review. 

While SDP proves useful in practice, a fundamental issue is to
numerically solve it. Theoretically, SDP is a convex conic problem  admitting polynomial-time algorithms, and for general SDPs, the interior point method (IPM) is known as a robust and efficient approach. Since the 1990s, high-performance SDP softwares based on the IPM have been developed, including {{\texttt{DSDP}}} {\cite{benson2008algorithm}}, \text{{\ttfamily{COPT}}}
{\cite{copt}}, \text{{\ttfamily{Mosek}}} {\cite{aps2019mosek}}, \text{{\ttfamily{Sedumi}}}
{\cite{polik2007sedumi}}, \text{{\ttfamily{SDPT3}}} {\cite{toh2012implementation}},
\text{{\ttfamily{CSDP}}} {\cite{borchers2006csdp}}, \text{{\ttfamily{SDPA}}} 
{\cite{yamashita2012latest}} and so forth. While most SDP codes are IPM-based, there are also  successful attempts using other methods (for example, {\cite{kocvara2006pensdp, kwasniewiczimplementation, yang2015sdpnal}})
and a list of these softwares is reviewed by {\cite{majumdar2020recent}}.

SDP solvers based on IPM variants exhibit competitive convergence behavior both theoretically and practically. Most softwares implement primal-dual path-following approaches using either infeasible start
{\cite{potra1998superlinearly}} or the embedding technique
{\cite{potra1998homogeneous}}, with {{\texttt{DSDP}}} being an exception:
{{\texttt{DSDP}}} adopts a dual IPM method based on the potential reduction framework developed by {\cite{benson1999mixed}}.
Since its initial release {\cite{benson2000solving}}, {{\texttt{DSDP}}} has gone through several major updates and evolved into a popular SDP software {\cite{benson2008algorithm}}.
To further improve {{\texttt{DSDP}}}, we make an important extension by incorporating the homogeneous self-dual embedding technique into the dual algorithm. This new implementation, named {{\texttt{HDSDP}}}, is presented in this manuscript.

The manuscript is organized as follows. \textbf{Section \ref{sec2}} 
describes the formulation of interest and some basic notations;
{\textbf{Section \ref{sec3}} reviews the dual-scaling algorithm
and describes how embedding technique can be applied. 
In {\textbf{Section \ref{sec4}} and \textbf{\ref{sec5}}, we introduce the practical aspects for {{\texttt{HDSDP}}}, including software design and algorithm configuration.  Last we present comprehensive numerical results on SDP problems.

\section{Formulation and Notations} \label{sec2}

{{\texttt{HDSDP}}} solves both primal and dual SDPs in standard form

{\center{\quad\qquad\qquad $\begin{array}{lccl}
  (P) & \min_{\mathbf{X}}  & \left\langle \mathbf{C}, \mathbf{X} \right\rangle & \\
  & \text{subject to} & \left\langle \mathbf{A}_i, \mathbf{X} \right\rangle = \mathbf{b}, & i = 1,
  \ldots, m\\
  &  & \mathbf{X} \in \mathbb{S}_+^n & 
\end{array}$\qquad$\begin{array}{lcc}
  (D) & \max_{\mathbf{y}, \mathbf{S}} & \mathbf{b}^{\top} \mathbf{y}\\
  & \text{subject to} & \sum_{i = 1}^m \mathbf{A}_i y_i + \mathbf{S} = \mathbf{C}\\
  &  & \mathbf{S} \in \mathbb{S}_+^n,
\end{array}$}}\\

where the problem data $\{\mathbf{A}_i\}, \mathbf{C}$ are $n \times n$ symmetric matrices
($\mathbb{S}^n$) and $\mathbf{b} \in \mathbb{R}^m$ is a real vector. Matrix inner
product $\langle \cdot, \cdot \rangle$ is defined by $\left\langle \mathbf{A}, \mathbf{X} \right\rangle := \sum_{i, j}
c_{i j} x_{i j}$ and $\mathbb{S}_+^n$ denotes the cone of positive
semidefinite matrices. For brevity we use $\mathbf{X} \succeq \textbf{0}$ to denote the
relation $\mathbf{X} \in \mathbb{S}_+^n$, and the linear map $\mathcal{A} : \mathbb{S}^n
\rightarrow \mathbb{R}^m$ together with its adjoint $\mathcal{A}^{\ast} : \mathbb{R}^m
\rightarrow \mathbb{S}^n$ is defined by $\mathcal{A} \mathbf{X} := \left(
\left\langle \mathbf{A}, \mathbf{X}_1 \right\rangle, \ldots, \left\langle \mathbf{A}_m, \mathbf{X}_m
\right\rangle \right)^{\top}$ and $\mathcal{A}^{\ast} \mathbf{y} := \sum_{i = 1}^m \mathbf{A}_i
y_i$. $\left\| \mathbf{A} \right\|_F := \sqrt{\sum_{i j} a_{i j}^2}$ denotes the
matrix Frobenius norm. With the above notations, we rewrite the primal and
dual problems by

{\center{\qquad\qquad\qquad\qquad$\begin{array}{lccl}
  (P) & \min_{\mathbf{X}}  & \left\langle \mathbf{C}, \mathbf{X} \right\rangle & \\
  & \text{subject to} & \mathcal{A} \mathbf{X} = \mathbf{b}, & \\
  &  & \mathbf{X} \succeq \textbf{0} & 
\end{array}$\qquad$\begin{array}{lcc}
  (D) & \max_{\mathbf{y}, \mathbf{S}} & \mathbf{b}^{\top} \mathbf{y}\\
  & \text{subject to} & \mathcal{A}^{\ast} \mathbf{y} + \mathbf{S} = \mathbf{C}\\
  &  & \mathbf{S} \succeq \textbf{0} .
\end{array}$}}\\

and primal and dual feasible regions are
\[\mathcal{F} (P) := \left\{ \mathbf{X} : \mathcal{A} \mathbf{X} = \mathbf{b}, \mathbf{X} \succeq \textbf{0} \right\} \qquad \mathcal{F} (D) := \left\{ \left( \mathbf{y}, \mathbf{S} \right) : \mathcal{A}^{\ast} \mathbf{y} +
\mathbf{S} = \mathbf{C}, \mathbf{S} \succeq \textbf{0} \right\}.\]
$\mathbf{X} \in \mathcal{F} (P)$ is primal
feasible and $\left( \mathbf{y}, \mathbf{S} \right) \in \mathcal{F} (D)$ is dual feasible.
The interior of these regions are denoted by $\mathcal{F}^0 (P),
\mathcal{F}^0 (D)$ and a solution in $\mathcal{F}^0$ is an interior
point solution.  {{\texttt{HDSDP}}} implements a dual method solving ({{\em P\/}}) and ({{\em D\/}}) through the self-dual embedding technique.

\section{Dual and Homogeneous Dual Scaling Algorithm }\label{sec3}

In this section, we briefly review the dual-scaling algorithm in
{{\texttt{DSDP}}}, and give its interpretation through Newton's method on the KKT system. Then we show how it naturally generalizes to the embedding.
\subsection{Dual-scaling Algorithm}
Dual-scaling method
{\cite{benson1999mixed}} works under three conditions. {\textbf{1)}} the
 data $\left\{ \mathbf{A}_i \right\}$ are linearly independent.
{\textbf{2)}} Both $(P)$ and $(D)$ admit an interior point
solution. {\textbf{3)}} an interior dual feasible solution $\left( \mathbf{y}^0,
\mathbf{S}^0 \right) \in \mathcal{F}^0 (D)$ is known. The first two conditions imply
strong duality and the existence of an optimal primal-dual solution pair $\left(
\mathbf{X}^{\ast}, \mathbf{y}^{\ast}, \mathbf{S}^{\ast} \right)$ satisfying complementarity
$\left\langle \mathbf{X}^{\ast}, \mathbf{S}^{\ast} \right\rangle = 0$. Also, the central path $\mathcal{C} (\mu) :=
\left\{ \left( \mathbf{X}, \mathbf{y}, \mathbf{S} \right) \in \mathcal{F}^0 (P) \times \mathcal{F}^0
(D) : \mathbf{X} \mathbf{S} = \mu \mathbf{I} \right\}$, which serves as a foundation of
path-following approaches, is well-defined. Given a centrality parameter $\mu$, 
dual method starts from
$\left( \mathbf{y}, \mathbf{S} \right) \in \mathcal{F}^0(D)$ and takes Newton's step towards the solution of the perturbed KKT system $\mathcal{A} \mathbf{X} = \mathbf{b},
\mathcal{A}^{\ast} \mathbf{y} + \mathbf{S} = \mathbf{C}$ and $\mathbf{X} \mathbf{S} = \mu \mathbf{I}$
\begin{align}\label{dsdpnewton}
  \mathcal{A} \left( \mathbf{X} + \Delta \mathbf{X} \right) & = \mathbf{b} \nonumber \\
  \mathcal{A}^{\ast} \Delta \mathbf{y} + \Delta \mathbf{S} & = \textbf{0} \\
  \mu \mathbf{S}^{- 1} \Delta \mathbf{S} \mathbf{S}^{- 1} + \Delta \mathbf{X} & = \mu \mathbf{S}^{- 1} - \mathbf{X} \nonumber, 
\end{align}
\revised{where the last relation linearizes $\mathbf{X} = \mu \mathbf{S}^{- 1}$ instead of $\mathbf{X} \mathbf{S} =
\mu \mathbf{I}$ and $\mathbf{S}^{- 1}$ is known as a scaling matrix. 
By geometrically driving $\mu$ to 0, dual-scaling eliminates (primal) infeasibility, approaches optimality, and finally solves the problem to some $\varepsilon$-optimal solution $(\mathbf{y_\varepsilon}, \mathbf{S_\varepsilon}) \in \mathcal{F}^0(D)$ such that $ \langle \mathbf{b}, \mathbf{y_\varepsilon} \rangle \geq  \langle \mathbf{b}, \mathbf{y^*} \rangle - \varepsilon$. Theory of dual potential reduction shows  an $\varepsilon$-optimal solution can be obtained in $\mathcal{O}(\log(1/\varepsilon))$ iterations ignoring dimension dependent constants.}

An important feature of dual-scaling is that $\mathbf{X}$ and $\Delta \mathbf{X}$
vanish in the Schur complement of \eqref{dsdpnewton}

\begin{eqnarray} \label{dsdpM}
	\left(\begin{array}{ccc}
     \left\langle \mathbf{A}_1, \mathbf{S}^{- 1} \mathbf{A}_1 \mathbf{S}^{- 1} \right\rangle & \cdots &
     \left\langle \mathbf{A}_1, \mathbf{S}^{- 1} \mathbf{A}_m \mathbf{S}^{- 1} \right\rangle\\
     \vdots & \ddots & \vdots\\
     \left\langle \mathbf{A}_m, \mathbf{S}^{- 1} \mathbf{A}_1 \mathbf{S}^{- 1} \right\rangle & \cdots &
     \left\langle \mathbf{A}_m, \mathbf{S}^{- 1} \mathbf{A}_m \mathbf{S}^{- 1} \right\rangle
   \end{array}\right) \Delta \mathbf{y} = \frac{1}{\mu} \mathbf{b} - \mathcal{A} \mathbf{S}^{- 1},
\end{eqnarray}\\
and the dual algorithm thereby avoids explicit reference to the primal
variable $\mathbf{X}$. For brevity we denote the left-hand side matrix in \eqref{dsdpM} by $\mathbf{M}$.
If $\left\{ \mathbf{A}_i \right\}, \mathbf{C}$ are sparse, any feasible dual slack $\mathbf{S} = \mathbf{C}
-\mathcal{A}^{\ast} \mathbf{y}$ inherits the aggregated sparsity pattern from the data and it is
computationally cheaper to iterate in the dual space.
Another feature of dual-scaling is the availability of primal
solution by solving a projection subproblem at the end of the algorithm \cite{benson2008algorithm}. The
aforementioned properties characterize the behavior of the dual-scaling method.

However, the desirable theoretical properties of the dual method is not free.  First, an initial dual feasible solution is needed, while obtaining such a
solution is often as difficult as solving the original problem. Second, due to
a lack of information from primal space, dual-scaling method has to be properly guided to avoid deviating from the central path. 
To overcome the aforementioned difficulties, {{\texttt{DSDP}}}
introduces artificial variables with big-$M$ penalty to ensure a nonempty interior and a trivial dual feasible solution. Moreover, a potential function is introduced 
to guide the dual iterations. This works well in practice and enhance performance of {{\texttt{DSDP}}}. We refer the interested readers to
{\cite{benson2000solving, benson2008algorithm}} for more implementation details.

\revised{Although {{\texttt{DSDP}}} proves efficient in real practice, the big-$M$
method requires prior estimation of the optimal solution to avoid losing optimality. 
Also, a large penalty often leads to numerical difficulties and
mis-classification of infeasible problems when the problem is ill-conditioned. 
Therefore it is natural to seek better alternatives to the big-$M$ method, and the self-dual embedding is an ideal candidate.
}
\subsection{Dual-scaling Algorithm using Embedding Technique}
Given a standard form SDP, its homogeneous self-dual model is a skew symmetric system containing the original problem data, whose non-trivial interior point solution certificates primal-dual feasibility. {\texttt{HDSDP}} adopts the following simplified embedding {\cite{potra1998homogeneous}}.
\begin{align}
  \mathcal{A} \mathbf{X} - \mathbf{b} \tau ={} & \textbf{0} \nonumber\\
  - \mathcal{A}^{\ast} \mathbf{y} + \mathbf{C} \tau - \mathbf{S} ={} & \textbf{0}  \\
  \mathbf{b}^{\top} \mathbf{y} - \langle \mathbf{C}, \mathbf{X} \rangle - \kappa ={}& 0 \nonumber\\
  \mathbf{X}, \mathbf{S} \succeq 0, \kappa, \tau  \geq{} & 0 \nonumber,
\end{align}
where $\kappa, \tau$ are homogenizing variables for infeasibility
detection. The central path of barrier parameter $\mu$ is given by
\begin{align}
  \mathcal{A} \mathbf{X} - \mathbf{b} \tau ={} & \textbf{0} \nonumber \\
  - \mathcal{A}^{\ast} \mathbf{y} + \mathbf{C} \tau - \mathbf{S} ={} & \textbf{0} \nonumber \\
  \mathbf{b}^{\top} \mathbf{y} - \langle \mathbf{C}, \mathbf{X} \rangle - \kappa ={} & 0 \nonumber \\
  \mathbf{X} \mathbf{S} = \mu \mathbf{I}, \kappa \tau ={} & \mu \label{hdsdpcompl} \\
  \mathbf{X}, \mathbf{S} \succeq 0, \kappa, \tau  \geq{} & 0. \nonumber
\end{align}
Here $\left( \mathbf{y}, \mathbf{S}, \tau \right)$ are jointly considered as dual variables.
Given an (infeasible) dual point $\left( \mathbf{y}, \mathbf{S}, \tau \right)$ such that $- \mathcal{A}^{\ast} \mathbf{y} +
\mathbf{C} \tau - \mathbf{S} = \mathbf{R}$, {{\texttt{HDSDP}}} selects a damping factor $\gamma \in
[0, 1]$ and takes Newton's step towards
\begin{align*}
  \mathcal{A} ( \mathbf{X} + \Delta \mathbf{X} ) - \mathbf{b} (\tau + \Delta \tau) & =\textbf{0}\\
  -\mathcal{A}^{\ast} ( \mathbf{y} + \Delta \mathbf{y}) + \mathbf{C} (\tau + \Delta \tau)
  - ( \mathbf{S} + \Delta \mathbf{S} ) & = - \gamma \mathbf{R}\\
  \mu \mathbf{S}^{- 1} \Delta \mathbf{S} \mathbf{S}^{- 1} + \Delta \mathbf{X} & = \mu \mathbf{S}^{- 1} - \mathbf{X},\\
  \mu \tau^{- 2} \Delta \tau + \Delta \kappa & =\mu \tau^{- 1} - \kappa,
\end{align*}
where, similar to {{\texttt{DSDP}}}, we modify \eqref{hdsdpcompl} and linearize $\mathbf{X} = \mu \mathbf{S}^{- 1}, \kappa
= \mu \tau^{- 1}$. We use this damping factor $\gamma$ and the barrier parameter $\mu$ to manage a trade-off between dual infeasibility, centrality and optimality. 
If we set $\gamma = 0$, then the Newton's direction $\left( \Delta \mathbf{y}, \Delta \tau \right)$ is computed through
the following Schur complement.

\begin{footnotesize}{
\begin{eqnarray*}
  \Bigg(\begin{array}{cc}
    \mu \mathbf{M} & - \mathbf{b} - \mu \mathcal{A} \mathbf{S}^{- 1} \mathbf{C} \mathbf{S}^{-1}\\
    - \mathbf{b} + \mu \mathcal{A} \mathbf{S}^{- 1} \mathbf{C} \mathbf{S}^{- 1} & - \mu (
    \langle \mathbf{C}, \mathbf{S}^{- 1} \mathbf{C} \mathbf{S}^{- 1} \rangle + \tau^{- 2} )
  \end{array}\Bigg) \left(\begin{array}{c}
    \Delta \mathbf{y}\\
    \Delta \tau
  \end{array}\right) =  \left(\begin{array}{c}
    \mathbf{b} \tau\\
    \mathbf{b}^{\top} \mathbf{y} - \mu \tau^{-1}
  \end{array}\right) - \mu \left(\begin{array}{c}
    \mathcal{A} \mathbf{S}^{- 1}\\
    \langle \mathbf{C}, \mathbf{S}^{- 1} \rangle
  \end{array}\right) + \mu \left(\begin{array}{c}
    \mathcal{A} \mathbf{S}^{- 1} \mathbf{R} \mathbf{S}^{- 1}\\
    \langle \mathbf{C}, \mathbf{S}^{- 1} \mathbf{R} \mathbf{S}^{- 1} \rangle
  \end{array}\right) \nonumber
\end{eqnarray*}}\end{footnotesize}In practice, {{\texttt{HDSDP}}} solves $\Delta \mathbf{y}_1 := \mathbf{M}^{- 1} \mathbf{b},
\Delta \mathbf{y}_2 := \mathbf{M}^{- 1} \mathcal{A} \mathbf{S}^{- 1}, \Delta \mathbf{y}_3 := \mathbf{M}^{- 1} \mathcal{A}
\mathbf{S}^{- 1} \mathbf{R} \mathbf{S}^{- 1}, \Delta \mathbf{y}_4 = \mathbf{M}^{- 1} \mathcal{A} \mathbf{S}^{- 1} \mathbf{C} \mathbf{S}^{-
1}$, plugs the solution into $\Delta \mathbf{y} = \frac{\tau + \Delta \tau}{\mu} \Delta \mathbf{y}_1 - \Delta
\mathbf{y}_2 + \Delta \mathbf{y}_3 + \Delta \mathbf{y}_4 \Delta \tau$ to get $\Delta \tau$, and finally assembles $\Delta \mathbf{y}$. With the above relations, $\mathbf{X} (\mu) := \mu \mathbf{S}^{- 1}
( \mathbf{S} - \mathbf{R} +\mathcal{A}^{\ast} \Delta \mathbf{y} - \mathbf{C} \Delta \tau ) \mathbf{S}^{-
1}$ satisfies $\mathcal{A} \mathbf{X} (\mu) - \mathbf{b} (\tau + \Delta \tau) = \textbf{0}$ and $\mathbf{X} (\mu)
\succeq \textbf{0}$ if and only if the backward Newton step $-\mathcal{A}^{\ast} ( \mathbf{y} - \Delta \mathbf{y} ) + \mathbf{C} (\tau
- \Delta \tau) - \mathbf{R} \succeq \textbf{0}$. When $-\mathcal{A}^{\ast} ( \mathbf{y} -
\Delta \mathbf{y} ) + \mathbf{C} (\tau - \Delta \tau) - \mathbf{R}$ is positive definite, a primal upper-bound follows from simple algebraic manipulation
\begin{align*}
\bar{z}  ={} & \langle \mathbf{C} \tau, \mathbf{X} (\mu) \rangle\\
	  ={} & \langle \mathbf{R} + \mathbf{S} + \mathcal{A}^{\ast} \mathbf{y}, \mu \mathbf{S}^{- 1} ( \mathbf{S} - \mathbf{R}
  +\mathcal{A}^{\ast} \Delta \mathbf{y} - \mathbf{C} \Delta \tau ) \mathbf{S}^{- 1}
  \rangle\\
  ={} & (\tau + \Delta \tau) \mathbf{b}^{\top} \mathbf{y} + n \mu + ( \mathcal{A} \mathbf{S}^{- 1} +
  \mathcal{A} \mathbf{S}^{- 1} \mathbf{R} \mathbf{S}^{- 1} )^{\top} \Delta \mathbf{y} + \mu (
  \langle \mathbf{C}, \mathbf{S}^{- 1} \rangle + \langle \mathbf{C}, \mathbf{S}^{- 1} \mathbf{C}
  \mathbf{S}^{- 1} \rangle ) \Delta \tau
\end{align*}
and dividing both sides by $\tau$. Alternatively, {{\texttt{HDSDP}}} extracts a primal
objective bound from the projection problem
\[   \min_{\mathbf{X}} ~ \| \mathbf{S}^{1 / 2} \mathbf{X} \mathbf{S}^{1 / 2} - \mu \mathbf{I} \|_F^2 \quad \text{subject to} \quad \mathcal{A} \mathbf{X} = \mathbf{b} \tau \]

whose optimal solution is $\mathbf{X}' (\mu) := \mu \mathbf{S}^{- 1} ( \mathbf{C} \tau -
\mathcal{A}^{\ast} ( \mathbf{y} - \Delta' \mathbf{y} ) - \mathbf{R} ) \mathbf{S}^{- 1}$. Here
$\Delta \mathbf{y}' = \frac{\tau}{\mu} \Delta \mathbf{y}_1 - \Delta \mathbf{y}_2$ and
\[ z' = \langle \mathbf{C} \tau, \mathbf{X}' (\mu) \rangle = \mu \{
   \langle \mathbf{R}, \mathbf{S}^{- 1} \rangle + ( \mathcal{A} \mathbf{S}^{- 1} \mathbf{R}
   \mathbf{S}^{- 1} + \mathcal{A} \mathbf{S}^{- 1})^{\top} \Delta' \mathbf{y} + n \} + \tau
   \mathbf{b}^{\top} \mathbf{y} . \]

Using the Newton's direction $\Delta \mathbf{y}$, {{\texttt{HDSDP}}} applies a simple ratio test
\begin{equation}
\alpha = \max \left\{ \alpha \in [0, 1] : \mathbf{S} + \alpha \Delta \mathbf{S} \succeq \textbf{0},
\tau + \alpha \Delta \tau \geq 0 \right\} \label{ratiotest}	
\end{equation}
 through a Lanczos procedure
{\cite{toh2002note}}. \revised{But to determine the aforementioned damping factor $\gamma$, we  
resort to the following adaptive heuristic: assuming that $\mu \rightarrow \infty, \tau = 1$ and fixing $\Delta\tau = 0$, the dual update can be decomposed by 
\begin{align}
  \mathbf{S} + \alpha \Delta \mathbf{S} ={} & \mathbf{S} + \alpha (\gamma
  \mathbf{R} -\mathcal{A}^{\ast} \Delta \mathbf{y}) \nonumber\\
  ={} & \mathbf{S} + \alpha (\gamma \mathbf{R} -\mathcal{A}^{\ast} (\gamma
  \Delta \mathbf{y}_3 - \Delta \mathbf{y}_2)) \nonumber\\
  ={} & \mathbf{S} + \alpha \mathcal{A}^{\ast} \Delta \mathbf{y}_2 + \alpha
  \gamma (\mathbf{R} -\mathcal{A}^{\ast} \Delta \mathbf{y}_3), \nonumber
\end{align}
where the first term $\mathbf{S} + \alpha \mathcal{A}^{\ast} \Delta \mathbf{y}_2$ is independent of $\gamma$ and improves centrality of the iteration. {\texttt{HDSDP}} conducts a Lanczos line-search to find $\alpha_c$ such that the logarithmic barrier function is improved \[- \log \det ( \mathbf{S}
+ \alpha_c \Delta \mathbf{S} ) - \log \det (\tau + \alpha_c \Delta \tau) \leq -
\log \det  \mathbf{S} - \log \det \tau. \] 
Then given $\alpha = \alpha_c$, by a second Lanczos line-search we find the maximum possible $\gamma \leq 1$ such that $\mathbf{S} + \alpha_c \Delta \mathbf{S} = \mathbf{S} + \alpha_c \mathcal{A}^{\ast} \Delta \mathbf{y}_2 + \alpha_c \gamma (\mathbf{R} -\mathcal{A}^{\ast} \Delta \mathbf{y}_3) \succeq 0$.
Finally,  a full Newton's direction is determined by $\gamma$ and
a third Lanczos procedure finally carries out the ratio test \eqref{ratiotest}.
The intuition of the above heuristic is to eliminate infeasibility under centrality restrictions, so that the iterations stay away from the boundary of the cone. After the ratio test, {\texttt{HDSDP}} updates $\mathbf{y} \leftarrow \mathbf{y} + 0.95 \alpha
\Delta \mathbf{y}$ and $\tau \leftarrow \tau + 0.95 \alpha \Delta \tau$ to ensure the feasibility of the next iteration. 
To make further use of the Schur complement $\mathbf{M}$, the above procedure is repeated several times in an iteration with $\mathbf{M}$ unchanged.}

\revised{
In {\texttt{HDSDP}}, the barrier parameter $\mu$ also significantly affects the algorithm. At the end of each iteration, {\texttt{HDSDP}} updates the
barrier parameter $\mu$ by $(z - \mathbf{b}^{\top} \mathbf{y} + \theta \left\|
\mathbf{R} \right\|_F ) / \rho n$, where $z$ is the best primal bound so far and  $\rho, \theta$ are pre-defined
parameters. By default $\rho = 4$ and $\theta = 10^8$. Heuristics also adjust $\mu$ based on $\alpha_c$ and $\gamma$. To get the best of both worlds, {\texttt{HDSDP}} implements the same dual-scaling algorithm as in \text{{\ttfamily{DSDP5.8}}}.  If dual infeasibility satisfies
$\left\| \mathcal{A}^{\ast} \mathbf{y} + \mathbf{S} - \mathbf{C} \right\|_F \leq \varepsilon \tau$ and $\mu$ is sufficiently small, {{\texttt{HDSDP}}} fixes $\tau = 1$, re-starts with $( \mathbf{y} / \tau, \mathbf{S} / \tau )$ and instead applies dual potential function to guide convergence. To sum up,
{{\texttt{HDSDP}}} implements strategies and computational tricks tailored for
the embedding and can switch to {{\texttt{DSDP5.8}}} once a dual feasible solution is available.}

\section{Initialization and Feasibility Certificate} \label{sec4}

Internally {{\texttt{HDSDP}}} deals with the following problem
\begin{eqnarray*}
  \min_{\mathbf{X}} & \left\langle \mathbf{C}, \mathbf{X} \right\rangle + \mathbf{u}^{\top} \mathbf{x}_u +
  \mathbf{l}^{\top} \mathbf{x}_l & \\
  \text{subject to} & \mathcal{A} \mathbf{X} + \mathbf{x}^u - \mathbf{x}^l = \mathbf{b} & \\
  & \mathbf{X} \succeq \textbf{0}, \mathbf{x}_u \geq \textbf{0}, \mathbf{x}_l \geq \textbf{0}. & 
\end{eqnarray*}

Together with its dual, the embedding is given by
\begin{align*}
  \mathcal{A} \mathbf{X} + \mathbf{x}^u - \mathbf{x}^l - \mathbf{b} \tau & = \textbf{0}\\
  - \mathcal{A}^{\ast} \mathbf{y} + \mathbf{C} \tau - \mathbf{S} & = \textbf{0}\\
  \mathbf{b}^{\top} \mathbf{y} - \left\langle \mathbf{C}, \mathbf{X} \right\rangle - \mathbf{u}^{\top} \mathbf{x}_u -
  \mathbf{l}^{\top} \mathbf{x}_l - \kappa & = 0\\
  \mathbf{X}, \mathbf{S} \succeq 0, &  \kappa, \tau \geq 0,
\end{align*}

\revised{where the primal problem is relaxed by two slack variables with penalty
$\textbf{{\textbf{l}}}, \mathbf{u}$ to prevent $\mathbf{y}$ going too large. Using the embedding, 
{{\texttt{HDSDP}}} needs no big-$M$ initialization
in the dual variable and starts from arbitrary $\mathbf{S} \succ \textbf{0}, \tau > 0$. By default {\texttt{HDSDP}} starts from $\tau = 1, \mathbf{y} = \mathbf{0}$ and $\mathbf{S}$ is initialized with $\mathbf{C} + 10^p\cdot\|\mathbf{C}\| \mathbf{I}$. One feature of the embedding is its capability to detect infeasibility.} Given infeasibility tolerance $\varepsilon_f$, {{\texttt{HDSDP}}} classifies the problem as primal unbounded, dual infeasible if $\left\| \mathbf{R}
\right\|_F > \varepsilon_f \tau, \tau / \kappa < \varepsilon_f$ and $\mu /
\mu_0 \leq \varepsilon_f^2$. If $\mathbf{b}^{\top} \mathbf{y} >
\varepsilon_f^{- 1}$, {{\texttt{HDSDP}}} starts checking if the Newton's
step $( \Delta \mathbf{y}, \Delta \mathbf{S} )$ is a dual improving ray. Once a ray is detected, the problem is classified as primal
infeasible and dual unbounded.

\section{HDSDP Software}\label{sec5}

\revised{{{\texttt{HDSDP}}} is written in \text{{\ttfamily{ANSI C}}} according to the standard of commercial numerical softwares}, and provides a
self-contained user interface. While {{\texttt{DSDP5.8}}} serves as a sub-routine library, 
{{\texttt{HDSDP}}} is designed to be a stand-alone SDP solver and re-written to accommodate the 
new computational techniques and third-party packages. After the user inputs the data and invokes the optimization routine, {{\texttt{HDSDP}}} goes through several modules
including input check, pre-solving, two-phase optimization and post-solving. The modules are implemented independently and are sequentially organized in a pipeline. \revised{Compared with  {{\texttt{DSDP}}}, two most important computational improvements of {{\texttt{HDSDP}}} lie in its conic KKT solver and its abstract linear system interface. To our knowledge, {{\texttt{HDSDP}}} has the most advanced KKT solver in the history of dual-scaling SDP softwares.
}

\subsection{Pre-solver}

One new feature of {{\texttt{HDSDP}}} is a special pre-solving module
 designed for SDPs. It inherits techniques from {{\texttt{DSDP5.8}}} and adds new strategies to work jointly with the optimization module. After the pre-solver is invoked, it first goes through $\{ \mathbf{A}_i \}$ two rounds to detect the possible low-rank structure. The first round uses Gaussian elimination for rank-one structure and the
second round applies eigenvalue decomposition from \text{{\ttfamily{Lapack}}}. Two
exceptions are when the data matrix is too dense or sparse. Unlike in {{\texttt{DSDP5.8}}} where eigen-decomposition is mandatory, matrices that are too dense to be eigen-decomposed efficiently will be skipped in {{\texttt{HDSDP}}}; if a matrix has very few entries, then a strategy from {{\texttt{DSDP5.8}}} is applied: {\textbf{1)}} a permutation gathers the non-zeros to a much smaller dense block. {\textbf{2)}} dense eigen routines from \text{{\ttfamily{Lapack}}} applies. {\textbf{3)}} the inverse permutation recovers the decomposition.

After detecting the hidden low-rank structures, the pre-solver moves on to collecting the sparsity and rank information of the matrices. The information will be kept for the KKT analysis to be described in \textbf{Section \ref{sec:kktsolver}}.


Finally, the pre-solver scales down the large objective coefficients by their Frobenius norm and goes on to identify the following seven structures. {\textbf{1)}}. Implied trace: constraints imply
$\ensuremath{\operatorname{tr}} ( \mathbf{X} ) = \theta$. {\textbf{2)}}. Implied dual bound:
constraints imply $\mathbf{l} \leq \mathbf{y} \leq \mathbf{u}$. {\textbf{3)}}. Empty
primal interior: constraints imply $\ensuremath{\operatorname{tr}} ( \mathbf{X} \mathbf{a} \mathbf{a}^{\top}
) \approx 0$. {\textbf{4)}}. Empty dual interior: constraints imply
$\mathcal{A}^{*} \mathbf{y} = \mathbf{C}$. {\textbf{5)}}. Feasibility problem: $\mathbf{C} = \textbf{0}$. {\textbf{6)}}. Dense problem: whether all the constraint matrices are totally dense.
{\textbf{7)}}. Multi-block problem: whether there are many SDP cones. For each of the cases the solver adjusts its internal parameters to enhance its numerical stability and convergence.

\subsection{Two-phase Optimization}

Underlie the procedure control of {{\texttt{HDSDP}}} is a two-phase method which integrates the
embedding technique (\text{{\ttfamily{Phase A}}}) and dual-scaling (\text{{\ttfamily{Phase B}}}). \
\text{{\ttfamily{Phase A}}} targets feasibility certificate, while \text{{\ttfamily{Phase B}}} aims to efficiently drive a dual-feasible solution to optimality.

\begin{figure}[h]
\centering
\includegraphics[scale=0.45]{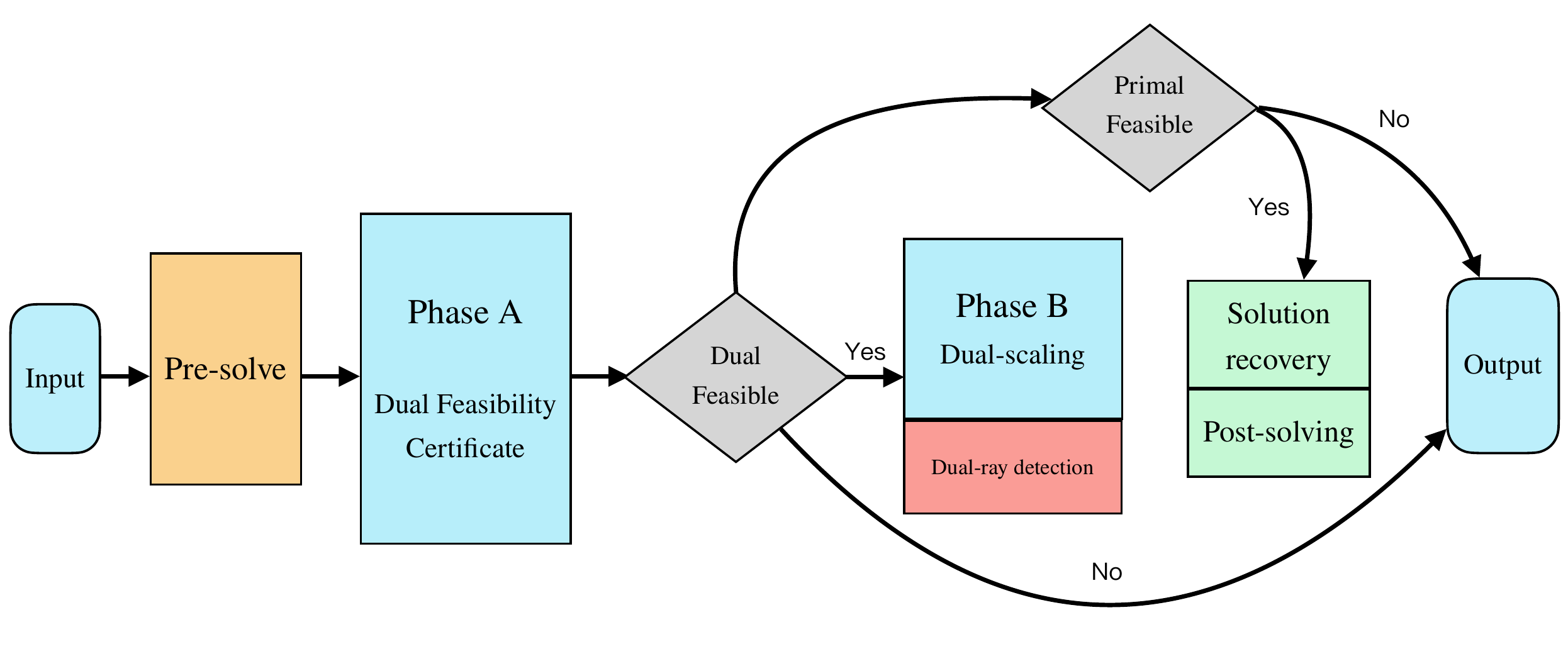}
  \caption{Pipeline of {{\texttt{HDSDP}}}}
\end{figure}

The two phases share the same backend computation routines but are
associated with different goals and strategies. {{\texttt{HDSDP}}} decides
which strategy to use based on the behavior of algorithm iterations.

\subsection{Conic KKT Solver} \label{sec:kktsolver}
\revised{
The most computationally intensive part in {{\texttt{HDSDP}}} is in the Schur complement matrix $\mathbf{M}$. In real-life applications, the SDP objective $\mathbf{C}$ often does not share the same sparsity or low-rank property as $\{ \mathbf{A}_i \}$. Therefore, the additional computation from $\mathcal{A} \mathbf{S}^{- 1} \mathbf{C} \mathbf{S}^{- 1}$ and $\langle \mathbf{C}, \mathbf{S}^{- 1} \mathbf{C} \mathbf{S}^{- 1} \rangle$ demands more efficient techniques to set up $\mathbf{M}$. }

\revised{Efficiency of setting up $\mathbf{M}$ depends mainly on two aspects: \textbf{1)}. the structure of the cones participating in forming the Schur complement, and \textbf{2)} the structure of the coefficients inside each cone. To exploit both structures, KKT solver in {{\texttt{HDSDP}}} implements two abstract interfaces, one from conic coefficient to conic operations, the other from cone to $\mathbf{M}$.}

\revised{The first interface allows {{\texttt{HDSDP}}} to set up $\mathbf{M}$ faster, while the second allows the extension of dual-scaling to arbitrary cones. These two interfaces already appeared in {{\texttt{DSDP5.8}}}, but unifying them in a conic KKT solver makes the algorithm implementation more compact. Till the date when this manuscript is written, four cones and five types SDP coefficients have been implemented by {{\texttt{HDSDP}}}.}

\revised{The interface from SDP coefficients to conic operations is directly relevant to the set up of $\mathbf{M}$. Using rank and sparsity information from the pre-solver, {{\texttt{HDSDP}}} analyzes the Schur system and generates a permutation $\sigma (m)$ of the rows of $\mathbf{M}$.
The idea is to permute the most computationally-expensive matrices to the left-most position, and this is a direct extension of {\cite{fujisawa1997exploiting}}. }

\[ \left(\begin{array}{ccc}
     \langle \mathbf{A}_{\sigma (1)}, \mathbf{S}^{- 1} \mathbf{A}_{\sigma (1)} \mathbf{S}^{- 1}
     \rangle & \cdots & \langle \mathbf{A}_{\sigma (1)}, \mathbf{S}^{- 1}
     \mathbf{A}_{\sigma (m)} \mathbf{S}^{- 1} \rangle\\
     & \ddots & \vdots\\
     &  & \langle \mathbf{A}_{\sigma (m)}, \mathbf{S}^{- 1} \mathbf{A}_{\sigma (m)} \mathbf{S}^{- 1}
     \rangle
   \end{array}\right) \]\\
\revised{After deciding the permutation, for each row of the permuted system, {{\texttt{HDSDP}}} chooses one out of a pool of five candidate techniques. We denote them by {\textbf{M1}} to {\textbf{M5}}, and they are efficient under different coefficient structures: Technique {\textbf{M1}} and {\textbf{M2}} are inherited from
{{\texttt{DSDP}}}. They exploit the low-rank structure of the coefficient data and
an eigen-decomposition of the data
$\mathbf{A}_i = \sum_{r = 1}^{\ensuremath{\operatorname{rank}} ( \mathbf{A}_i )} \lambda_{i r}
   \mathbf{a}_{i, r} \mathbf{a}^{\top}_{i, r}$
needs to be computed at the beginning of the algorithm; Technique {\textbf{M3}}, {\textbf{M4}} and {\textbf{M5}} exploit sparsity and need evaluation of $\mathbf{S}^{-1}$ {\cite{fujisawa1997exploiting}} every iteration.\\}

{\begin{minipage}{0.8\textwidth}
		{\textbf{KKT Technique M1}}
\begin{enumerate}[leftmargin=25pt]
  \item {\textbf{Setup}} $\mathbf{B}_{\sigma (i)} = \sum_{r =
  1}^{\ensuremath{\operatorname{rank}} ( \mathbf{A}_{\sigma (i)} )} \lambda_r \big( \mathbf{S}^{- 1}
   \mathbf{a}_{\sigma (i), r} \big) ( \mathbf{S}^{- 1} \mathbf{a}_{\sigma (i), r})^{\top}$
  
  \item {\textbf{Compute}} $M_{\sigma (i) \sigma (j)} = \langle
  \mathbf{B}_{\sigma (i)}, \mathbf{A}_{\sigma (j)} \rangle, \text{for all } j \geq i$
\end{enumerate}
{\textbf{KKT Technique M2}}
\begin{enumerate}[leftmargin=25pt]
  \item {\textbf{Setup}} $\mathbf{S}^{- 1} \mathbf{a}_{\sigma (i), r}, r = 1, \ldots,
  r_{\sigma (i)}$
  \item {\textbf{Compute}} $M_{\sigma (i) \sigma (j)} =  \sum_{r =
  1}^{\ensuremath{\operatorname{rank}} ( \mathbf{A}_{\sigma (i)} )} \lambda_r ( \mathbf{S}^{- 1}
  \mathbf{a}_{\sigma (i), r} )^{\top} \mathbf{A}_{\sigma (j)} ( \mathbf{S}^{- 1}
  \mathbf{a}_{\sigma (i), r} )$
\end{enumerate}
{\textbf{KKT Technique M3}}
\begin{enumerate}[leftmargin=25pt]
  \item {\textbf{Setup}} $\mathbf{B}_{\sigma (i)} = \mathbf{S}^{- 1} \mathbf{A}_{\sigma
  (i)} \mathbf{S}^{- 1}$
  \item {\textbf{Compute}} $M_{\sigma (i) \sigma (j)} = \langle
  \mathbf{B}_{\sigma (i)}, \mathbf{A}_{\sigma (j)} \rangle, \text{for all } j \geq i$
\end{enumerate}

{\textbf{KKT Technique M4}}
\begin{enumerate}[leftmargin=25pt]
  \item {\textbf{Setup}} $\mathbf{B}_{\sigma (i)} = \mathbf{S}^{- 1} \mathbf{A}_{\sigma
  (i)}$
  
  \item {\textbf{Compute}} $M_{\sigma (i) \sigma (j)} = \langle
  \mathbf{B}_{\sigma (i)} \mathbf{S}^{- 1}, \mathbf{A}_{\sigma (j)} \rangle, \text{for all } j
  \geq i$
\end{enumerate}
{\textbf{KKT Technique M5}}
\begin{enumerate}[leftmargin=25pt]
  \item {\textbf{Compute}} $M_{\sigma (i) \sigma (j)} = \langle \mathbf{S}^{-
  1} \mathbf{A}_{\sigma (i)} \mathbf{S}^{- 1}, \mathbf{A}_{\sigma (j)} \rangle, \text{for all } j \geq
  i$ directly
\end{enumerate}
\end{minipage}}

\begin{table}[h]
  \centering
  \caption{Approximate flops for each strategy. $f_i$ is the number of nonzeros of $\mathbf{A}_i$; $r_i$ is rank of $\mathbf{A}_i$, $\kappa$ is the slow down ratio of discontinuous memory access}
  \begin{tabular}{ccc}
  \toprule
    KKT Technique & Floating point operations & Extra cost \\
    \midrule
    {\textbf{M1}} & $r_{\sigma (i)} (n^2 + 2 n^2) + \kappa \sum_{j \geq i}
    f_{\sigma (j)}$ & 0\\
    {\textbf{M2}} & $r_{\sigma (i)} ( n^2 + \kappa \sum_{j \geq i}
    f_{\sigma (j)} )$ & 0 \\
    {\textbf{M3}} & $n \kappa f_{\sigma (i)} + n^3 + \sum_{j \geq i} \kappa
    f_{\sigma (j)}$ & $n^3$ \\
    {\textbf{M4}} & $n \kappa f_{\sigma (i)} + \sum_{j \geq i} \kappa (n +
    1) f_{\sigma (j)}$ & $n^3$\\
    {\textbf{M5}} & $\kappa (2 \kappa f_{\sigma (i)} + 1)  \sum_{j \geq i}
    f_{\sigma (j)}$ & $n^3$\\
    \bottomrule
  \end{tabular}
\end{table}

\revised{This newly developed KKT solver improves the speed of {{\texttt{HDSDP}}} on a broad set of instances. And to our knowledge {{\texttt{HDSDP}}} is the first SDP software that simultaneously incorporates all the commonly-used Schur complement strategies.}

\subsection{Linear System Interface and Parallel Computation}

\revised{
Linear system solving is the core of any interior point software, and most of the linear system solvers nowadays provide support for multi-threading. In {{\texttt{HDSDP}}}, matrix decomposition is 
required for both the dual matrix $\mathbf{S}$ and the Schur complement $\mathbf{M}$, both of which would need different decomposition routines based on matrix data structure and numerical conditioning. To meet this requirement, {{\texttt{HDSDP}}} adopts a plug-in type abstract linear system interface which allows users to adopt any linear system routine. \textbf{Table \ref{table:linsys}} summarizes the routines supported by {{\texttt{HDSDP}}}, and by default the multi-threaded Intel MKL routines are used and threading is controlled by a user-specified parameter.
\begin{table}[h]
\centering
  \caption{Linear systems in {{\texttt{HDSDP}}}} \label{table:linsys}
  \begin{tabular}{cccc}
      \toprule
    Linear system type & Implementation & Parallel & Available in Public \\
      \midrule
    Dense Restarted PCG & {{\texttt{HDSDP}}} & Yes & Yes \\
    Positive definite sparse direct & (MKL) Pardiso \cite{wang2014intel} & Yes & No\\
    Quasi-definite sparse direct & \texttt{COPT} \cite{copt} & Yes & No\\
    Positive-definite dense direct & \texttt{COPT} \cite{copt} & Yes & No\\
    Positive definite sparse direct & SuiteSparse \cite{davis2006direct} & No & Yes\\    
    Positive definite dense direct  & (MKL) Lapack \cite{wang2014intel} & Yes & Yes\\
    Symmetric indefinite dense direct & (MKL) Lapack \cite{wang2014intel} & Yes & Yes\\
    \bottomrule
  \end{tabular}
\end{table}}

\revised{Aside from the third-party routines, {{\texttt{HDSDP}}} itself implements a pre-conditioned conjugate gradient (CG) method with restart to solve the Schur complement system. The maximum number of iterations is
chosen around $50 / m$ and is heuristically adjusted. Either the diagonal of
$\mathbf{M}$ or its Cholesky decomposition is chosen as pre-conditioner and
after a Cholesky pre-conditioner is computed, {{\texttt{HDSDP}}} reuses it for 
the consecutive solves till a heuristic determines that the current
pre-conditioner is outdated. When the algorithm approaches optimality, $\mathbf{M}$
might become ill-conditioned and {{\texttt{HDSDP}}} switches to {\textsf{LDL}} decomposition in case Cholesky fails.}

\section{Computational results}

\revised{
The efficiency of {{\texttt{DSDP5.8}}} has been proven through
years of computational experience, and {{\texttt{HDSDP}}} aims to achieve
further improvement on instances where dual method has
advantage over the primal-dual method.  In this section, we introduce several types of SDPs suitable for the dual method and we compare the performance of {{\texttt{HDSDP}}}, {{\texttt{DSDP5.8}}} and \text{{\ttfamily{COPT 6.5}}} (fastest solver on Mittelmann's benchmark) on several benchmark datasets to verify the performance improvement of
{{\texttt{HDSDP}}}.
}

\subsection{Experiment Setup}

\revised{
We configure the experiment as follows
\begin{enumerate}[leftmargin=*]
	\item (Testing platform). The tests are run on an \text{{\ttfamily{intel i11700K}}} PC with 128GB memory and 12 threads.We choose \texttt{HDSDP}, \texttt{DSDP5.8} and \texttt{COPT6.5} are selected as benchmark softwares.
	\item (Compiler and third-party dependency). Both \texttt{HDSDP} and \texttt{DSDP5.8} are compiled using \texttt{icc} with \texttt{-O2} optimization and linked with threaded version intel MKL. Executable of \texttt{COPT} is directly obtained from its official website.
	\item (Threading). We set the number of MKL threads to be 12 for \texttt{DSDP5.8}; the \texttt{Threads} parameter of \texttt{COPT} is also set to 12. To enhance reproducibility \texttt{HDSDP} uses 8 threads.
	\item (Dataset). Datasets are chosen from three sources: 1). SDP benchmark datasets from Hans Mittelmann's website \cite{mittelmann2003independent}; 2). SDP benchmark datasets from \texttt{SDPLIB} \cite{borchers1999sdplib}. 3). Optimal diagonal pre-conditioner SDPs generated according to \cite{qu2020diagonal}.
	\item (Algorithm). Default methods in \texttt{HDSDP}, \texttt{DSDP5.8}, and the primal-dual interior point method implemented in \texttt{COPT} are compared.
	\item (Tolerance). All the feasibility and optimality tolerances are set to $5\times10^{-6}$.
	\item (Solution status). We adopt the broadly accepted DIMACs error to determine if a solution is qualified. According to \cite{mittelmann2003independent}, if any of the DIMACS errors exceeds $10^{-2}$, the solution is considered invalid.
	\item (Performance Metric). We use shifted geometric mean to compare the overall speed between different solvers.
\end{enumerate}
}

\subsection{Maximum Cut}

The SDP relaxation of the max-cut problem is represented by
\begin{eqnarray*}
  \min_{\mathbf{X}} & \left\langle \mathbf{C}, \mathbf{X} \right\rangle & \\
  \text{subject to} & \ensuremath{\operatorname{diag}} \left( \mathbf{X} \right) = \textbf{1} & \\
  & \mathbf{X} \succeq \textbf{0} . & 
\end{eqnarray*}
Let $\mathbf{e}_i$ be the $i$-th column of the identity matrix and the constraint
$\ensuremath{\operatorname{diag}} \left( \mathbf{X} \right) = \mathbf{e}$ is decomposed into $\left\langle \mathbf{X}, \mathbf{e}_i
\mathbf{e}_i^{\top} \right\rangle = 1, i = 1, \ldots, n$. Note that $\mathbf{e}_i \mathbf{e}_i^{\top}$
is rank-one and has only one non-zero entry, {\textbf{M2}} and
{\textbf{M5}} can greatly reduce the computation of the Schur matrix.

\begin{table}[h]
  \caption{Max-cut problems}
  \centering
  \begin{tabular}{crrrcrrr}
    \toprule
    Instance & {{\texttt{HDSDP}}} & {{\texttt{DSDP5.8}}} & \text{{\ttfamily{COPT
    v6.5}}} & Instance & {{\texttt{HDSDP}}} & {{\texttt{DSDP5.8}}} &
    \text{{\ttfamily{COPT v6.5}}}\\
    \midrule
    \text{{\ttfamily{mcp100}}} & {0.03} & {0.02} & 0.11 & \text{{\ttfamily{maxG51}}} & {1.45} &
    2.52 & 14.40 \\
    \text{{\ttfamily{mcp124-1}}} & 0.04 & {0.02} & 0.17 & \text{{\ttfamily{maxG55}}} & {38.21}
    & 273.60 & 1096.02 \\
    \text{{\ttfamily{mcp124-2}}} & 0.04 & {0.02} & 0.17 & \text{{\ttfamily{maxG60}}} & {87.59}
    & 535.20 & 2926.11 \\
    \text{{\ttfamily{mcp124-3}}} & 0.05 & {0.02} & 0.16 & \text{{\ttfamily{G40\_mb}}} & {15.77}
    & 8.77 & 98.97\\
    \text{{\ttfamily{mcp124-4}}} & 0.06 & {0.05} & 0.17 & \text{{\ttfamily{G40\_mc}}} & {6.53} &
    18.68 & 80.45\\
    \text{{\ttfamily{mcp250-1}}} & 0.09 & {0.08} & 0.70 & \text{{\ttfamily{G48\_mb}}} & {17.83}
    & 12.48 & 183.95\\
    \text{{\ttfamily{mcp250-2}}} & {0.08} & 0.09 & 0.72 & \text{{\ttfamily{G48mc}}} & {5.09} &
    8.43 & 118.79\\
    \text{{\ttfamily{mcp250-3}}} & {0.09} & 0.12 & 0.65 & \text{{\ttfamily{G55mc}}} & {38.06} &
    168.1 & 1026.10\\
    \text{{\ttfamily{mcp250-4}}} & {0.19} & 0.16 & 0.71 & \text{{\ttfamily{G59mc}}} & {48.73} &
    302.3 & 1131.24\\
    \text{{\ttfamily{mcp500-1}}} & 0.26 & {0.21} & 2.78 & \text{{\ttfamily{G60\_mb}}} & {211.22}
    & 213.4 & 5188.11\\
    \text{{\ttfamily{mcp500-2}}} & {0.25} & 0.32 & 2.82 & \text{{\ttfamily{G60mc}}} & {208.96} &
    218.8 & 5189.29 \\
    \text{{\ttfamily{mcp500-3}}} & {0.30} & 0.50 & 2.59 & \text{{\ttfamily{torusg3-8}}} & {0.50}
    & 0.77 & 1.04\\
    \text{{\ttfamily{mcp500-4}}} & {0.39} & 0.86 & 2.86 & \text{{\ttfamily{torusg3-15}}} &
    {15.30} & 178.8 & 137.60\\
    \text{{\ttfamily{maxG11}}} & 0.60 & {0.46} & 7.15 & \text{{\ttfamily{toruspm3-8-50}}} &
    {0.41} & 0.40 & 2.65\\
    \text{{\ttfamily{maxG32}}} & {4.68} & 3.66 & 68.32 & \text{{\ttfamily{toruspm3-15-50}}} &
    {13.32} & 43.67 & 309.25 \\
    \bottomrule
  \end{tabular}
\end{table}

Computational experience suggests that on large-scale sparse max-cut instances, {{\texttt{HDSDP}}} 
is more than $5$ times faster than {{\texttt{DSDP5.8}}}. Two exceptions are \text{{\ttfamily{G60mc}}} and \text{{\ttfamily{G60\_mb}}}, where the aggregated sparsity pattern of the dual matrix is lost due to the dense objective coefficient, and thereby the dense MKL routines take up most of the computation.

\subsection{Graph Partitioning}

The SDP relaxation of the graph partitioning problem is given by
\begin{eqnarray*}
  \min_{\mathbf{X}} & \langle \mathbf{C}, \mathbf{X} \rangle & \\
  \text{subject to} & \ensuremath{\operatorname{diag}} \left( \mathbf{X} \right) = \textbf{1} & \\
  & \langle \textbf{{\textbf{1}}1}^{\top}, \mathbf{X} \rangle = \beta &
  \\
  & k \mathbf{X} - \textbf{{\textbf{1}}1}^{\top} \succeq \textbf{0} & \\
  & \mathbf{X} \geq \textbf{0}, & 
\end{eqnarray*}
where $\textbf{1}$ denotes the all-one vector and $k, \beta$ are the problem
parameters. Although the dual $\mathbf{S}$ no longer enjoys sparsity, the low-rank
structure is still available to accelerate convergence.

\begin{table}[h]
  \caption{Graph partitioning problems}
  \begin{tabular}{crrrcrrr}
    \toprule
    Instance & {{\texttt{HDSDP}}} & {{\texttt{DSDP5.8}}} & \text{{\ttfamily{COPT
    v6.5}}} & Instance & {{\texttt{HDSDP}}} & {{\texttt{DSDP5.8}}} &
    \text{{\ttfamily{COPT v6.5}}}\\
    \midrule
    \text{{\ttfamily{gpp100}}} & {0.05} & 0.05 & 0.19 & \text{{\ttfamily{gpp250-4}}} & 0.15 &
    {0.14} & 1.52\\
    \text{{\ttfamily{gpp124-1}}} & {0.07} & 0.07 & 0.33 & \text{{\ttfamily{gpp500-1}}} & {0.89}
    & 0.44 & 5.52\\
    \text{{\ttfamily{gpp124-2}}} & {0.06} & Failed & 0.30 & \text{{\ttfamily{gpp500-2}}} & 0.80
    & {0.43} & 5.43\\
    \text{{\ttfamily{gpp124-3}}} & {0.06} & {0.06} & 0.31 & \text{{\ttfamily{gpp500-3}}} & 0.65
    & {0.36} & 5.54\\
    \text{{\ttfamily{gpp124-4}}} & 0.06 & {0.06} & 0.29 & \text{{\ttfamily{gpp500-4}}} & {0.68}
    & {0.38} & 5.40\\
    \text{{\ttfamily{gpp250-1}}} & {0.20} & Failed & 1.65 & \text{{\ttfamily{bm1}}} & {2.28} &
    1.18 & 18.45 \\
    \text{{\ttfamily{gpp250-2}}} & {0.17} & 0.14 & 1.21 & \text{{\ttfamily{biomedP}}} & {171.64}
    & \text{{\ttfamily{Failed}}} & \text{{\ttfamily{Failed}}}\\
    \text{{\ttfamily{gpp250-3}}} & 0.17 & {0.13} & 1.41 & \\
    \bottomrule
  \end{tabular}
\end{table}
On graph partitioning instances, we see that {{\texttt{HDSDP}}} has comparable performance to {{\texttt{DSDP}}} but is more robust on some of the problems.

\subsection{Optimal Diagonal Pre-conditioning}

The optimal diagonal pre-conditioning problem originates from
{\cite{qu2020diagonal}}, where given a matrix $\mathbf{B} \succ \textbf{0}$, finding a
diagonal matrix $\mathbf{D}$ to minimize the condition number $\kappa (
\mathbf{D}^{- 1 / 2} \mathbf{B} \mathbf{D}^{- 1 / 2} )$ can be modeled as an
SDP. The formulation for optimal diagonal pre-conditioning is given by
\begin{eqnarray*}
  \max_{\tau, \mathbf{D}} & \tau & \\
  \text{subject to} & \mathbf{D} \preceq \mathbf{B} & \\
  & \tau \mathbf{B} - \mathbf{D} \preceq \textbf{0} . & 
\end{eqnarray*}
Expressing $\mathbf{D} = \sum_i \mathbf{e}_i \mathbf{e}_i^{\top} d_i$, the problem is exactly
in the SDP dual form. If $\mathbf{B}$ is also sparse, the problem can be
efficiently solved using the dual method.
\begin{table}[H]
\centering
  \caption{Optimal diagonal pre-conditioning problems}
  \begin{tabular}{rrrr}
    \toprule
    Instance & {{\texttt{HDSDP}}} & {{\texttt{DSDP5.8}}} & \text{{\ttfamily{COPT
    v6.5}}}\\
    \midrule
    \text{{\ttfamily{diag-bench-1000-0.01}}} & {37.670} & 207.500 & 38.61\\
    \text{{\ttfamily{diag-bench-2000-0.05}}} & 276.960 & 971.700 & {161.17} \\
    \text{{\ttfamily{diag-bench-west0989}}} & {35.280} & \text{{\ttfamily{Failed}}} & 76.900 \\
    \text{{\ttfamily{diag-bench-DK01R}}} & {5.010} & \text{{\ttfamily{Failed}}}  & \text{{\ttfamily{Failed}}} \\
    \text{{\ttfamily{diag-bench-micromass\_10NN}}} & 20.510  &  38.45 & {17.430} \\
    \bottomrule
  \end{tabular}
\end{table}


\subsection{Other Problems}

So far {{\texttt{HDSDP}}} is tested and tuned over a broad set of benchmarks
including \text{{\ttfamily{SDPLIB}}} {\cite{borchers1999sdplib}} and Hans
Mittelmann's sparse SDP benchmark {\cite{mittelmann2003independent}}. 
Using geometric mean as the metric, compared to {{\texttt{DSDP5.8}}}, {{\texttt{HDSDP}}} achieves a speedup of 21\% on \text{{\ttfamily{SDPLIB}}} and around 17\% speedup on Hans Mittelmann's benchmark dataset. Below we list some example benchmark datasets of nice structure for {{\texttt{HDSDP}}} to exploit.
%

\begin{table}[h]
\centering
  \caption{Features of several benchmark problems}
  \begin{tabular}{cccrrr}
    \toprule
    Instance & Background & Feature & {{\texttt{HDSDP}}} &
    {{\texttt{DSDP5.8}}} & \text{{\ttfamily{COPT v6.5}}}\\
    \midrule
    \text{{\ttfamily{checker\_1.5}}} & unknown & sparse, low-rank & {39.55} & 64.04 &
    868.37\\
    \text{{\ttfamily{foot}}} & unknown & sparse, low-rank & {26.68} & 13.65 & 245.41\\
    \text{{\ttfamily{hand}}} & unknown & low-rank & {6.60} & 2.64 & 49.39\\
    \text{{\ttfamily{ice\_2.0}}} & unknown & low-rank & {284.90} & 504.70 & 8561.74\\
    \text{{\ttfamily{p\_auss2\_3.0}}} & unknown & sparse, low-rank & {528.00} & 1066.00
    & 1111.72\\
    \text{{\ttfamily{rendl1\_2000\_1e-6}}} & unknown & low-rank & {16.17} & 14.38 &
    111.43\\
    \text{{\ttfamily{trto4}}} & topology design & sparse, low-rank & {6.25} & 6.30 &
    27.16\\
    \text{{\ttfamily{trto5}}} & topology design & sparse, low-rank & {67.22} & 75.20 &
    391.83\\
    \text{{\ttfamily{sensor\_500b}}} & sensor network localization & sparse, low-rank
    & 19.84 & {35.11} & {8.97}\\
    \text{{\ttfamily{sensor\_1000b}}} & sensor network localization & sparse,
    low-rank & 76.39 & {98.18} & {38.96}\\
    \bottomrule
  \end{tabular}
\end{table}


\section{When (not) to use DSDP/HDSDP}

While {{\texttt{HDSDP}}} is designed for general SDPs, it targets the problems
more tractable in the dual form than by the primal-dual methods. This is
the principle for the techniques implemented by {{\texttt{HDSDP}}}.
Here are some rules in mind when deciding whether to use the dual method (or
{{\texttt{HDSDP}}}).
\begin{enumerate}[leftmargin=20pt]
  \item Does the problem enjoys nice dual structure?
  
  Many combinatorial problems have formulations friendly to the dual methods.
  Some typical features include (aggregated) sparsity and low-rank structure.
  Dual methods effectively exploit these features by iterating in the dual space and
  using efficient computational tricks. If the problem is dense and most
  constraints are full-rank, dual method has no advantage over the primal-dual
  solvers due to {\textbf{1)}} comparable iteration cost to primal-dual
  methods. {\textbf{2)}} more iterations for convergence.
  
  \item Do we need the primal optimal solution or just the optimal value?
  
  For some applications dual method fails to recover a correct primal solution
  due to numerical difficulties. If the optimal value is sufficient, there is
  no problem. But if an accurate primal optimal solution is always necessary,
  it is better to be more careful and to test the recovery procedure in case
  of failure at the last step.
  
  \item Do we need to certificate infeasibility strictly?
  
  One weakness of the dual method is the difficulty in infeasibility
  certificate. Although on the dual side this issue is addressed by
  {{\texttt{HDSDP}}} using the embedding, dual methods still suffer from
  failure to identify primal infeasibility.
  
  \item Is dual-feasibility hard to attain?
  
  The first phase of {{\texttt{HDSDP}}} adopts the infeasible Newton's method
  and focuses on eliminating the dual infeasibility. This principle works well
  if the dual constraints are relatively easy to satisfy, but if this
  condition fails to hold (for example, empty dual interior), experiments suggest the embedding
  spend a long time deciding feasibility. In this case it is suggested using
  {{\texttt{DSDP5.8}}} or supply an initial dual solution.
\end{enumerate}
\section{Conclusions}

\revised{We propose an extension of the dual-scaling algorithm based on
the embedding technique. The resultant solver, \text{{\ttfamily{HDSDP}}}, is presented to
demonstrate how dual method can be effectively integrated with the embedding. 
\text{{\ttfamily{HDSDP}}} is developed in parallel to \text{{\ttfamily{DSDP5.8}}} and
is entailed with several newly added features, especially an advanced conic KKT solver. The solver
exhibits promising performance on several benchmark datasets and is under
active development. Now \text{{\ttfamily{HDSDP}}} works as one of the candidate SDP methods in the state-of-the-art commercial solver \texttt{COPT}}. Users are welcome to try the solver and provide valuable suggestions.
\section{Acknowledgement}

We thank Dr. Qi Huangfu and Dr. Joachim Dahl from \text{{\ttfamily{COPT}}} development team \cite{copt} for their valuable suggestions in the solver design and implementation. We also appreciate
Hans Mittelmann's efforts in benchmarking the solver. Finally, we sincerely
respect the developers of \text{{\ttfamily{DSDP}}} for their precious suggestions
{\cite{benson2008algorithm}} and their invaluable efforts getting
\text{{\ttfamily{DSDP}}} through all along the way. It is the efficient and elegant
implementation from \text{{\ttfamily{DSDP5.8}}} that guides \text{{\ttfamily{HDSDP}}} to
where it is.

\bibliography{sdpref.bib}

\begin{thebibliography}{10}

\bibitem{aps2019mosek}
Mosek ApS.
\newblock Mosek optimization toolbox for matlab.
\newblock {\em User’s Guide and Reference Manual, Version}, 4, 2019.

\bibitem{benson2008algorithm}
Steven~J Benson and Yinyu Ye.
\newblock Algorithm 875: Dsdp5--software for semidefinite programming.
\newblock {\em ACM Transactions on Mathematical Software (TOMS)}, 34(3):1--20,
  2008.

\bibitem{benson2000solving}
Steven~J Benson, Yinyu Ye, and Xiong Zhang.
\newblock Solving large-scale sparse semidefinite programs for combinatorial
  optimization.
\newblock {\em SIAM Journal on Optimization}, 10(2):443--461, 2000.

\bibitem{benson1999mixed}
Steven~J Benson, Yinyu Yeb, and Xiong Zhang.
\newblock Mixed linear and semidefinite programming for combinatorial and
  quadratic optimization.
\newblock {\em Optimization Methods and Software}, 11(1-4):515--544, 1999.

\bibitem{biswas2004semidefinite}
Pratik Biswas and Yinyu Ye.
\newblock Semidefinite programming for ad hoc wireless sensor network
  localization.
\newblock In {\em Proceedings of the 3rd international symposium on Information
  processing in sensor networks}, pages 46--54, 2004.

\bibitem{borchers1999sdplib}
Brian Borchers.
\newblock Sdplib 1.2, a library of semidefinite programming test problems.
\newblock {\em Optimization Methods and Software}, 11(1-4):683--690, 1999.

\bibitem{borchers2006csdp}
Brian Borchers.
\newblock Csdp user’s guide, 2006.

\bibitem{davis2006direct}
Timothy~A Davis.
\newblock {\em Direct methods for sparse linear systems}.
\newblock SIAM, 2006.

\bibitem{fujisawa1997exploiting}
Katsuki Fujisawa, Masakazu Kojima, and Kazuhide Nakata.
\newblock Exploiting sparsity in primal-dual interior-point methods for
  semidefinite programming.
\newblock {\em Mathematical Programming}, 79(1):235--253, 1997.

\bibitem{copt}
Dongdong Ge, Qi~Huangfu, Zizhuo Wang, Jian Wu, and Yinyu Ye.
\newblock Cardinal optimizer (copt) user guide.
\newblock {\em arXiv preprint arXiv:2208.14314}, 2022.

\bibitem{goemans1995improved}
Michel~X Goemans and David~P Williamson.
\newblock Improved approximation algorithms for maximum cut and satisfiability
  problems using semidefinite programming.
\newblock {\em Journal of the ACM (JACM)}, 42(6):1115--1145, 1995.

\bibitem{hayashi2016quantum}
Masahito Hayashi.
\newblock {\em Quantum information theory}.
\newblock Springer, 2016.

\bibitem{kocvara2006pensdp}
Michal Kocvara, Michael Stingl, and PENOPT GbR.
\newblock Pensdp users guide (version 2.2).
\newblock {\em PENOPT GbR}, 1435:1436, 2006.

\bibitem{kwasniewiczimplementation}
Tomasz Kwasniewicz and Fran{\c{c}}ois Glineur.
\newblock Implementation of a semidefinite optimization solver in the julia
  programming language.
\newblock 2021.

\bibitem{laurent2009sums}
Monique Laurent.
\newblock Sums of squares, moment matrices and optimization over polynomials.
\newblock In {\em Emerging applications of algebraic geometry}, pages 157--270.
  Springer, 2009.

\bibitem{laurent2005semidefinite}
Monique Laurent and Franz Rendl.
\newblock Semidefinite programming and integer programming.
\newblock {\em Handbooks in Operations Research and Management Science},
  12:393--514, 2005.

\bibitem{majumdar2020recent}
Anirudha Majumdar, Georgina Hall, and Amir~Ali Ahmadi.
\newblock Recent scalability improvements for semidefinite programming with
  applications in machine learning, control, and robotics.
\newblock {\em Annual Review of Control, Robotics, and Autonomous Systems},
  3:331--360, 2020.

\bibitem{mittelmann2003independent}
Hans~D Mittelmann.
\newblock An independent benchmarking of sdp and socp solvers.
\newblock {\em Mathematical Programming}, 95(2):407--430, 2003.

\bibitem{polik2007sedumi}
Imre Polik, Tamas Terlaky, and Yuriy Zinchenko.
\newblock Sedumi: a package for conic optimization.
\newblock In {\em IMA workshop on Optimization and Control, Univ. Minnesota,
  Minneapolis}. Citeseer, 2007.

\bibitem{potra1998homogeneous}
Florian~A Potra and Rongqin Sheng.
\newblock On homogeneous interrior-point algorithms for semidefinite
  programming.
\newblock {\em Optimization Methods and Software}, 9(1-3):161--184, 1998.

\bibitem{potra1998superlinearly}
Florian~A Potra and Rongqin Sheng.
\newblock A superlinearly convergent primal-dual infeasible-interior-point
  algorithm for semidefinite programming.
\newblock {\em SIAM Journal on Optimization}, 8(4):1007--1028, 1998.

\bibitem{qu2020diagonal}
Zhaonan Qu, Yinyu Ye, and Zhengyuan Zhou.
\newblock Diagonal preconditioning: Theory and algorithms.
\newblock {\em arXiv preprint arXiv:2003.07545}, 2020.

\bibitem{so2007theory}
Anthony Man-Cho So and Yinyu Ye.
\newblock Theory of semidefinite programming for sensor network localization.
\newblock {\em Mathematical Programming}, 109(2):367--384, 2007.

\bibitem{toh2002note}
Kim-Chuan Toh.
\newblock A note on the calculation of step-lengths in interior-point methods
  for semidefinite programming.
\newblock {\em Computational Optimization and Applications}, 21(3):301--310,
  2002.

\bibitem{toh2012implementation}
Kim-Chuan Toh, Michael~J Todd, and Reha~H T{\"u}t{\"u}nc{\"u}.
\newblock On the implementation and usage of sdpt3--a matlab software package
  for semidefinite-quadratic-linear programming, version 4.0.
\newblock In {\em Handbook on semidefinite, conic and polynomial optimization},
  pages 715--754. Springer, 2012.

\bibitem{vandenberghe1996semidefinite}
Lieven Vandenberghe and Stephen Boyd.
\newblock Semidefinite programming.
\newblock {\em SIAM review}, 38(1):49--95, 1996.

\bibitem{wang2014intel}
Endong Wang, Qing Zhang, Bo~Shen, Guangyong Zhang, Xiaowei Lu, Qing Wu, Yajuan
  Wang, Endong Wang, Qing Zhang, Bo~Shen, et~al.
\newblock Intel math kernel library.
\newblock {\em High-Performance Computing on the Intel{\textregistered} Xeon
  Phi™: How to Fully Exploit MIC Architectures}, pages 167--188, 2014.

\bibitem{wolkowicz2005semidefinite}
Henry Wolkowicz.
\newblock Semidefinite and cone programming bibliography/comments.
\newblock {\em http://orion. uwaterloo. ca/\~{}
  hwolkowi/henry/book/fronthandbk. d/sdpbibliog. pdf}, 2005.

\bibitem{yamashita2012latest}
Makoto Yamashita, Katsuki Fujisawa, Mituhiro Fukuda, Kazuhiro Kobayashi,
  Kazuhide Nakata, and Maho Nakata.
\newblock Latest developments in the sdpa family for solving large-scale sdps.
\newblock In {\em Handbook on semidefinite, conic and polynomial optimization},
  pages 687--713. Springer, 2012.

\bibitem{yang2015sdpnal}
Liuqin Yang, Defeng Sun, and Kim-Chuan Toh.
\newblock Sdpnal++: a majorized semismooth newton-cg augmented lagrangian
  method for semidefinite programming with nonnegative constraints.
\newblock {\em Mathematical Programming Computation}, 7(3):331--366, 2015.

\end{thebibliography}
\bibliographystyle{plain}
\newpage
\appendix\section{Mittlelmann's Benchmark Test}

This section summarizes the benchmark result of \texttt{HDSDP} on part of Hans Mittelmann's SDP collection. \cite{mittelmann2003independent}. The main setup is consistent with the experiment section. If an instance fails, its solution time is recorded to be 40000 seconds.

\subsection{Summary Statistics}

\begin{table}[H]
  \centering
\caption{Summary statistics on the benchmark}
  \begin{tabular}{ccc}
  \toprule
  Solver & Number of solved instances & Shifted geometric mean \\
  \midrule
  \texttt{HDSDP} & 67 & 228.64 \\
    \texttt{DSDP5.8} & 62 & 276.96 \\
    \texttt{COPT v6.5} & 72 & 100.53 \\
  	\bottomrule
  \end{tabular}	
\end{table}

\subsection{Detailed Solution Statistics}
\begin{table}[H]
  \centering
\caption{Tables of detailed solution statistics}
  \begin{tabular}{cc}
  \toprule
  Solver & Table \\
  \midrule
   \texttt{HDSDP} & Table \ref{hans-hdsdp-1} and  \ref{hans-hdsdp-2}   \\
    \texttt{DSDP5.8} & Table \ref{hans-dsdp-1} and  \ref{hans-dsdp-2} \\
    \texttt{COPT v6.5} & Table \ref{hans-copt-1} and  \ref{hans-copt-2} \\
  	\bottomrule
  \end{tabular}	
\end{table}

{\small{\begin{table}[h]
  \begin{center}
    \caption{HDSDP: Mittelmann's Benchmark Test Part 1}
  \label{hans-hdsdp-1}
  \resizebox{\textwidth}{!}{%
  \begin{tabular}{cccccccc}
      \toprule
      Instance & Error 1 & Error 2 & Error 3 & Error 4 & Error 5 & Error 6 &
      Time\\
\midrule
            1dc.1024 & 4.16e-14 & 0.00e+00 & 0.00e+00 & -0.00e+00 & 5.56e-07 & 5.56e-07 & 730.950 \\ 
            1et.1024 & 2.90e-12 & 0.00e+00 & 0.00e+00 & -0.00e+00 & 8.20e-08 & 8.20e-08 & 62.870 \\ 
            1tc.1024 & 2.72e-13 & 0.00e+00 & 0.00e+00 & -0.00e+00 & 2.60e-07 & 2.60e-07 & 44.270 \\ 
            1zc.1024 & 2.79e-14 & 0.00e+00 & 0.00e+00 & -0.00e+00 & 1.30e-07 & 1.30e-07 & 369.410 \\ 
AlH\_1-Sigma+\_STO-6GN14r20g1T2\_5 & 2.20e-01 & -0.00e+00 & 0.00e+00 & -0.00e+00 & -9.95e-01 & 0.00e+00 & Failed \\ 
            Bex2\_1\_5 & 7.15e-10 & 0.00e+00 & 9.02e-12 & 0.00e+00 & 3.98e-07 & 1.15e-07 & 199.260 \\ 
BH2\_2A1\_STO-6GN7r14g1T2 & 1.40e-10 & 0.00e+00 & 2.37e-09 & 0.00e+00 & 1.36e-07 & 3.17e-07 & 282.880 \\ 
               biggs & 8.36e-10 & 2.29e-15 & 1.87e-12 & 0.00e+00 & 2.76e-08 & 2.95e-08 & 8.580 \\ 
           broyden25 & 9.98e-14 & 0.00e+00 & 0.00e+00 & -0.00e+00 & 1.92e-08 & 1.92e-08 & 1066.460 \\ 
         Bst\_jcbpaf2 & 1.15e-13 & 0.00e+00 & 1.66e-10 & 0.00e+00 & 1.94e-07 & 1.52e-07 & 317.920 \\ 
               buck4 & 2.27e-11 & 0.00e+00 & 0.00e+00 & -0.00e+00 & 3.83e-07 & 2.07e-07 & 21.100 \\ 
               buck5 & 1.39e-07 & 0.00e+00 & 0.00e+00 & -0.00e+00 & 6.48e-04 & 3.36e-04 & 185.080 \\ 
             butcher & 3.03e-07 & 0.00e+00 & 3.63e-07 & 0.00e+00 & -9.42e-05 & 8.56e-09 & 333.150 \\ 
          cancer\_100 & 2.85e-12 & 0.00e+00 & 0.00e+00 & -0.00e+00 & 3.58e-09 & 1.42e-08 & 213.620 \\ 
CH2\_1A1\_STO-6GN8r14g1T2 & 2.05e-11 & 0.00e+00 & 4.35e-09 & 0.00e+00 & 1.53e-07 & 2.65e-07 & 271.080 \\ 
         checker\_1.5 & 6.71e-14 & 0.00e+00 & 0.00e+00 & -0.00e+00 & 7.17e-07 & 7.17e-07 & 39.550 \\ 
            chs\_5000 & 0.00e+00 & 0.00e+00 & 0.00e+00 & -0.00e+00 & 3.34e-08 & 3.34e-08 & 57.980 \\ 
             cnhil10 & 3.39e-08 & 0.00e+00 & 0.00e+00 & -0.00e+00 & 7.39e-09 & 1.73e-08 & 37.520 \\ 
             cphil12 & 8.00e-15 & 0.00e+00 & 0.00e+00 & -0.00e+00 & 1.06e-09 & 1.06e-09 & 253.960 \\ 
       diamond\_patch & 1.00e+00 & -0.00e+00 & 5.25e-05 & 0.00e+00 & 1.00e+00 & 1.00e+00 & Failed \\ 
e\_moment\_quadknap\_17\_100\_0.5\_2\_2 & 1.63e-10 & 0.00e+00 & 4.92e-09 & 0.00e+00 & -1.71e-06 & 8.77e-09 & 118.180 \\ 
e\_moment\_stable\_17\_0.5\_2\_2 & 1.05e-12 & 0.00e+00 & 3.63e-07 & 0.00e+00 & -1.20e-05 & 2.76e-08 & 123.590 \\ 
                foot & 9.94e-06 & 0.00e+00 & 0.00e+00 & -0.00e+00 & 2.18e-04 & 1.20e-05 & 26.680 \\ 
              G40\_mb & 4.02e-09 & 0.00e+00 & 0.00e+00 & -0.00e+00 & 2.66e-07 & 2.17e-07 & 15.770 \\ 
               G40mc & 7.81e-15 & 0.00e+00 & 0.00e+00 & -0.00e+00 & 3.66e-07 & 3.66e-07 & 6.530 \\ 
              G48\_mb & 3.28e-05 & 0.00e+00 & 0.00e+00 & -0.00e+00 & -6.27e-04 & 4.31e-07 & 17.830 \\ 
               G48mc & 2.38e-14 & 0.00e+00 & 0.00e+00 & -0.00e+00 & 2.36e-07 & 2.36e-07 & 5.090 \\ 
               G55mc & 2.82e-15 & 0.00e+00 & 0.00e+00 & -0.00e+00 & 1.67e-07 & 1.67e-07 & 38.060 \\ 
               G59mc & 1.28e-14 & 0.00e+00 & 0.00e+00 & -0.00e+00 & 2.33e-07 & 2.33e-07 & 48.730 \\ 
              G60\_mb & 8.93e-09 & 0.00e+00 & 0.00e+00 & -0.00e+00 & 2.46e-07 & 2.52e-07 & 211.220 \\ 
               G60mc & 8.93e-09 & 0.00e+00 & 0.00e+00 & -0.00e+00 & 2.46e-07 & 2.52e-07 & 208.960 \\ 
H3O+\_1-A1\_STO-6GN10r16g1T2\_5 & 2.09e-12 & 0.00e+00 & 6.27e-12 & 0.00e+00 & 2.32e-07 & 2.32e-07 & 1086.470 \\ 
                hand & 1.82e-08 & 0.00e+00 & 0.00e+00 & -0.00e+00 & 6.54e-07 & 3.73e-07 & 6.600 \\ 
             ice\_2.0 & 1.48e-13 & 0.00e+00 & 0.00e+00 & -0.00e+00 & 1.72e-07 & 1.72e-07 & 284.900 \\ 
            inc\_1200 & 1.85e-04 & 0.00e+00 & 0.00e+00 & -0.00e+00 & -2.27e-03 & 2.53e-06 & 46.420 \\ 
           neosfbr25 & 8.64e-13 & 0.00e+00 & 0.00e+00 & -0.00e+00 & 7.00e-08 & 7.00e-08 & 543.240 \\ 
         neosfbr30e8 & 2.27e-11 & 0.00e+00 & 0.00e+00 & -0.00e+00 & 4.15e-08 & 4.15e-08 & 2733.590 \\ 
                neu1 & 1.66e-11 & 0.00e+00 & 1.60e-07 & 0.00e+00 & 4.85e-06 & 2.08e-05 & 86.940 \\ 
               neu1g & 1.92e-08 & 0.00e+00 & 7.94e-10 & 0.00e+00 & 4.09e-08 & 8.37e-08 & 65.970 \\ 
               neu2c & 5.24e-03 & 0.00e+00 & 3.79e-10 & 0.00e+00 & -9.87e-05 & 9.07e-04 & 183.370 \\ 
      \bottomrule
    \end{tabular}
  }  
\end{center}
\end{table}}}

{\small{\begin{table}[h]
\caption{HDSDP: Mittelmann's Benchmark Test Part 2}
\label{hans-hdsdp-2}
  \begin{center}
  \resizebox{\textwidth}{!}{%
  \begin{tabular}{cccccccc}
      \toprule
      Instance & Error 1 & Error 2 & Error 3 & Error 4 & Error 5 & Error 6 &
      Time\\
      \midrule
                neu2 & 5.44e-01 & 0.00e+00 & 6.14e-03 & 0.00e+00 & 9.98e-01 & 1.28e-01 & Failed \\ 
               neu2g & 3.76e-10 & 1.66e-16 & 7.94e-10 & 0.00e+00 & 3.39e-09 & 2.42e-08 & 57.240 \\ 
                neu3 & 1.11e-14 & 0.00e+00 & 6.42e-07 & 0.00e+00 & -1.97e-06 & 2.04e-08 & 788.670 \\ 
               neu3g & 2.76e-14 & 0.00e+00 & 0.00e+00 & -0.00e+00 & 7.71e-09 & 2.04e-08 & 861.030 \\ 
  NH2-.1A1.STO6G.r14 & 1.07e-12 & 0.00e+00 & 3.37e-09 & 0.00e+00 & 2.74e-07 & 3.25e-07 & 259.130 \\ 
NH3\_1-A1\_STO-6GN10r16g1T2\_5 & 4.14e-12 & 0.00e+00 & 6.20e-09 & 0.00e+00 & 1.29e-07 & 2.18e-07 & 1076.710 \\ 
 NH4+.1A1.STO6GN.r18 & 6.50e-10 & 0.00e+00 & 3.22e-09 & 0.00e+00 & 8.83e-06 & 9.02e-06 & 3601.580 \\ 
            nonc\_500 & 5.47e-11 & 0.00e+00 & 0.00e+00 & -0.00e+00 & 7.58e-08 & 7.58e-08 & 6.080 \\ 
         p\_auss2\_3.0 & 9.88e-15 & 0.00e+00 & 0.00e+00 & -0.00e+00 & 2.37e-07 & 2.37e-07 & 311.050 \\ 
               rabmo & 1.36e-11 & 0.00e+00 & 1.63e-07 & 0.00e+00 & -4.28e-05 & 1.18e-08 & 66.380 \\ 
             reimer5 & 4.65e-02 & -0.00e+00 & 1.15e-07 & 0.00e+00 & 9.38e-01 & 0.00e+00 & Failed \\ 
    rendl1\_2000\_1e-6 & 1.74e-11 & 0.00e+00 & 0.00e+00 & -0.00e+00 & 3.88e-07 & 3.87e-07 & 16.170 \\ 
            ros\_2000 & 4.44e-16 & 0.00e+00 & 0.00e+00 & -0.00e+00 & 9.09e-08 & 9.09e-08 & 9.110 \\ 
             ros\_500 & 4.52e-11 & 0.00e+00 & 0.00e+00 & -0.00e+00 & 1.33e-07 & 1.34e-07 & 2.040 \\ 
              rose15 & 1.13e-08 & 0.0e+00 & 1.40e-07 & 0.00e+00 & -9.69e-05 & 8.37e-09 & 46.850 \\ 
         sensor\_1000 & 1.06e-11 & 0.00e+00 & 0.00e+00 & -0.00e+00 & 3.19e-07 & 7.21e-08 & 98.180 \\ 
          sensor\_500 & 3.38e-12 & 0.00e+00 & 0.00e+00 & -0.00e+00 & 1.21e-07 & 2.27e-08 & 35.110 \\ 
              shmup4 & 1.83e-09 & 0.00e+00 & 0.00e+00 & -0.00e+00 & 6.63e-07 & 4.67e-07 & 73.280 \\ 
              shmup5 & 2.18e-08 & 0.00e+00 & 8.54e-11 & 0.00e+00 & -2.91e-06 & 4.98e-07 & 663.480 \\ 
           swissroll & 9.48e-06 & 0.00e+00 & 0.00e+00 & -0.00e+00 & -7.34e-03 & 1.19e-07 & 31.170 \\ 
              taha1a & 9.01e-11 & 0.00e+00 & 1.84e-07 & 0.00e+00 & -1.41e-07 & 1.78e-07 & 135.320 \\ 
              taha1b & 2.20e-11 & 0.00e+00 & 2.51e-10 & 0.00e+00 & 4.90e-08 & 6.07e-08 & 391.190 \\ 
              taha1c & 1.50e-10 & 0.00e+00 & 3.76e-07 & 0.00e+00 & -2.06e-07 & 1.93e-07 & 1631.100 \\ 
            theta102 & 1.89e-14 & 0.00e+00 & 0.00e+00 & -0.00e+00 & 6.65e-08 & 6.65e-08 & 1196.390 \\ 
             theta12 & 1.84e-14 & 0.00e+00 & 0.00e+00 & -0.00e+00 & 2.89e-08 & 2.89e-08 & 167.570 \\ 
       tiger\_texture & 4.33e-05 & 0.00e+00 & 1.05e-09 & 0.00e+00 & 6.48e-04 & 9.81e-04 & 47.520 \\ 
               trto4 & 2.88e-09 & 0.00e+00 & 0.00e+00 & -0.00e+00 & 2.49e-07 & 8.11e-08 & 6.250 \\ 
               trto5 & 3.26e-08 & 0.00e+00 & 0.00e+00 & -0.00e+00 & -3.72e-07 & 1.84e-07 & 67.220 \\ 
              vibra4 & 1.94e-06 & 0.00e+00 & 0.00e+00 & -0.00e+00 & 6.11e-06 & 3.21e-06 & 27.360 \\ 
              vibra5 & 4.31e-07 & 0.00e+00 & 1.46e-08 & 0.00e+00 & 2.11e-04 & 1.11e-04 & 293.290 \\ 
              yalsdp & 2.00e-07 & 0.00e+00 & 0.00e+00 & -0.00e+00 & 9.83e-07 & 9.82e-07 & 118.100 \\ 

      \bottomrule
    \end{tabular}
  }  
\end{center}  
\end{table}}}

{\small{\begin{table}[h]
  \caption{DSDP5.8: Mittelmann's Benchmark Test Part 1}
  \label{hans-dsdp-1}
  \begin{center}
  \resizebox{\textwidth}{!}{%
  \begin{tabular}{cccccccc}
      \toprule
      Instance & Error 1 & Error 2 & Error 3 & Error 4 & Error 5 & Error 6 &
      Time\\
\midrule
            1dc.1024 & 2.00e-06 & 0.00e+00 & 0.00e+00 & 0.00e+00 & 8.26e-01 & 7.95e+00 & Failed \\ 
            1et.1024 & 6.22e+01 & 0.00e+00 & 0.00e+00 & 0.00e+00 & 6.55e-01 & 1.04e+02 & Failed \\ 
            1tc.1024 & 5.13e-10 & 0.00e+00 & 0.00e+00 & 0.00e+00 & 6.34e-08 & 7.46e-08 & 44.310 \\ 
            1zc.1024 & 5.43e-10 & 0.00e+00 & 0.00e+00 & 0.00e+00 & 5.47e-09 & 7.61e-09 & 497.000 \\ 
AlH\_1-Sigma+\_STO-6GN14r20g1T2\_5 & 2.27e-10 & 0.00e+00 & 0.00e+00 & 0.00e+00 & 9.53e-05 & 1.09e-04 & 7296.000 \\ 
            Bex2\_1\_5 & 1.84e-06 & 0.00e+00 & 0.00e+00 & 0.00e+00 & 6.79e-07 & 1.15e-06 & 47.120 \\ 
BH2\_2A1\_STO-6GN7r14g1T2 & 3.16e-10 & 0.00e+00 & 0.00e+00 & 0.00e+00 & 1.16e-07 & 1.18e-07 & 182.200 \\ 
               biggs & 7.21e-09 & 0.00e+00 & 0.00e+00 & 0.00e+00 & 1.98e-09 & 4.08e-08 & 3.729 \\ 
           broyden25 & 5.14e-11 & 0.00e+00 & 0.00e+00 & 0.00e+00 & 1.20e-07 & 1.68e-07 & 392.900 \\ 
         Bst\_jcbpaf2 & 3.30e-08 & 0.00e+00 & 0.00e+00 & 0.00e+00 & 8.72e-08 & 2.12e-07 & 60.910 \\ 
               buck4 & 7.06e-06 & 0.00e+00 & 0.00e+00 & 0.00e+00 & 2.20e-07 & 3.54e-07 & 18.760 \\ 
               buck5 & 2.94e-05 & 0.00e+00 & 0.00e+00 & 0.00e+00 & 1.21e-06 & 1.59e-06 & 421.900 \\ 
             butcher & 8.33e-05 & 4.61e-14 & 0.00e+00 & 0.00e+00 & 3.03e-06 & 3.21e-06 & 125.200 \\ 
          cancer\_100 & 1.77e-09 & 0.00e+00 & 0.00e+00 & 0.00e+00 & 3.60e-08 & 4.68e-06 & 137.700 \\ 
CH2\_1A1\_STO-6GN8r14g1T2 & 1.61e-10 & 0.00e+00 & 0.00e+00 & 0.00e+00 & 1.34e-07 & 1.35e-07 & 184.800 \\ 
         checker\_1.5 & 3.19e-09 & 0.00e+00 & 0.00e+00 & 0.00e+00 & 5.25e-07 & 5.25e-07 & 64.040 \\ 
            chs\_5000 & 1.84e-10 & 0.00e+00 & 0.00e+00 & 0.00e+00 & 2.23e-07 & 6.53e-07 & 539.100 \\ 
             cnhil10 & 7.00e-07 & 0.00e+00 & 0.00e+00 & 0.00e+00 & 3.46e-06 & 1.56e-05 & 11.710 \\ 
             cphil12 & 1.03e-11 & 0.00e+00 & 0.00e+00 & 0.00e+00 & 2.05e-06 & 6.89e-06 & 85.760 \\ 
       diamond\_patch & 1.41e-07 & 0.00e+00 & 0.00e+00 & 0.00e+00 & 5.04e-04 & 8.43e-02 & Failed \\ 
e\_moment\_quadknap\_17\_100\_0.5\_2\_2 & 1.84e-08 & 2.61e-15 & 0.00e+00 & 0.00e+00 & 3.42e-09 & 1.20e-07 & 35.770 \\ 
e\_moment\_stable\_17\_0.5\_2\_2 & 1.68e-01 & 1.41e-16 & 0.00e+00 & 0.00e+00 & -1.83e-02 & 6.56e-01 & Failed \\ 
                foot & 2.90e-09 & 0.00e+00 & 0.00e+00 & 0.00e+00 & 1.08e-07 & 7.66e-08 & 13.650 \\ 
              G40\_mb & 2.00e-09 & 0.00e+00 & 0.00e+00 & 0.00e+00 & 6.49e-08 & 4.10e-06 & 8.774 \\ 
               G40mc & 2.26e-09 & 0.00e+00 & 0.00e+00 & 0.00e+00 & 1.10e-07 & 1.12e-07 & 18.680 \\ 
              G48\_mb & 3.05e-09 & 0.00e+00 & 0.00e+00 & 0.00e+00 & 4.18e-07 & 5.24e-06 & 12.480 \\ 
               G48mc & 2.77e-09 & 0.00e+00 & 0.00e+00 & 0.00e+00 & 1.23e-07 & 1.23e-07 & 8.434 \\ 
               G55mc & 3.57e-09 & 0.00e+00 & 0.00e+00 & 0.00e+00 & 5.15e-07 & 5.18e-07 & 168.100 \\ 
               G59mc & 3.57e-09 & 0.00e+00 & 0.00e+00 & 0.00e+00 & 3.43e-07 & 3.44e-07 & 302.300 \\ 
              G60\_mb & 7.10e-09 & 0.00e+00 & 0.00e+00 & 0.00e+00 & 2.30e-06 & 2.35e-05 & 213.400 \\ 
               G60mc & 7.10e-09 & 0.00e+00 & 0.00e+00 & 0.00e+00 & 2.30e-06 & 2.35e-05 & 218.800 \\ 
H3O+\_1-A1\_STO-6GN10r16g1T2\_5 & 3.94e-12 & 0.00e+00 & 0.00e+00 & 0.00e+00 & 1.17e-07 & 1.20e-07 & 821.800 \\ 
                hand & 1.38e-09 & 0.00e+00 & 0.00e+00 & 0.00e+00 & 8.10e-08 & 4.03e-07 & 2.641 \\ 
             ice\_2.0 & 4.67e-09 & 0.00e+00 & 0.00e+00 & 0.00e+00 & 5.12e-07 & 5.12e-07 & 504.700 \\ 
            inc\_1200 & 1.58e+00 & 0.00e+00 & 0.00e+00 & 0.00e+00 & 2.24e-01 & 2.13e-01 & Failed \\ 
             mater-5 & 5.19e-09 & 0.00e+00 & 0.00e+00 & 0.00e+00 & 5.66e-07 & 6.32e-07 & 20.280 \\ 
             mater-6 & 1.06e-08 & 0.00e+00 & 0.00e+00 & 0.00e+00 & 8.67e-07 & 9.51e-07 & 68.250 \\ 
           neosfbr25 & 1.30e-10 & 0.00e+00 & 0.00e+00 & 0.00e+00 & 1.52e-08 & 1.91e-08 & 233.800 \\ 
         neosfbr30e8 & 1.65e-10 & 0.00e+00 & 0.00e+00 & 0.00e+00 & 5.42e-09 & 6.60e-09 & 1286.000 \\ 
                neu1 & 7.00e-11 & 0.00e+00 & 0.00e+00 & 0.00e+00 & 2.35e-04 & 2.08e-02 & Failed \\ 
               neu1g & 8.17e-09 & 0.00e+00 & 0.00e+00 & 0.00e+00 & 2.25e-07 & 3.77e-05 & 38.390 \\ 
               neu2c & 1.84e-03 & 2.70e-15 & 0.00e+00 & 0.00e+00 & -2.00e-03 & 1.70e-05 & 88.330 \\ 
      \bottomrule
    \end{tabular}
  }  
\end{center}
\end{table}}}

{\small{\begin{table}[h]
\caption{DSDP5.8: Mittelmann's Benchmark Test Part 2}
\label{hans-dsdp-2}
  \begin{center}
  \resizebox{\textwidth}{!}{%
  \begin{tabular}{cccccccc}
      \toprule
      Instance & Error 1 & Error 2 & Error 3 & Error 4 & Error 5 & Error 6 &
      Time\\
      \midrule
      neu2 & 3.42e-13 & 0.00e+00 & 0.00e+00 & 0.00e+00 & 8.97e-08 & 1.28e-05 & 30.270 \\ 
               neu2g & 7.84e-08 & 0.00e+00 & 0.00e+00 & 0.00e+00 & -1.73e-09 & 7.55e-06 & 28.600 \\ 
                neu3 & 2.23e-09 & 0.00e+00 & 0.00e+00 & 0.00e+00 & 4.13e-08 & 5.76e-04 & 194.300 \\ 
               neu3g & 1.22e-08 & 0.00e+00 & 0.00e+00 & 0.00e+00 & 3.77e-08 & 4.41e-04 & 252.100 \\ 
  NH2-.1A1.STO6G.r14 & 2.56e-01 & 8.85e-16 & 0.00e+00 & 0.00e+00 & -5.24e-02 & 4.50e-02 & Failed \\ 
NH3\_1-A1\_STO-6GN10r16g1T2\_5 & 5.71e-12 & 0.00e+00 & 0.00e+00 & 0.00e+00 & 1.45e-07 & 1.48e-07 & 802.500 \\ 
 NH4+.1A1.STO6GN.r18 & 1.65e-01 & 9.75e-15 & 0.00e+00 & 0.00e+00 & -4.93e-02 & 2.94e-03 & Failed \\ 
            nonc\_500 & 1.40e-09 & 0.00e+00 & 0.00e+00 & 0.00e+00 & 2.38e-07 & 5.83e-07 & 2.225 \\ 
         p\_auss2\_3.0 & 4.83e-09 & 0.00e+00 & 0.00e+00 & 0.00e+00 & 1.96e-07 & 1.96e-07 & 528.000 \\ 
               rabmo & 6.39e-04 & 5.41e-16 & 0.00e+00 & 0.00e+00 & -7.59e-06 & 4.24e-07 & 37.470 \\ 
             reimer5 & 4.95e-08 & 4.57e-15 & 0.00e+00 & 0.00e+00 & 2.41e-09 & 1.24e-06 & 244.200 \\ 
    rendl1\_2000\_1e-6 & 1.91e-09 & 0.00e+00 & 0.00e+00 & 0.00e+00 & 3.53e-08 & 3.57e-08 & 14.380 \\ 
            ros\_2000 & 6.88e-11 & 0.00e+00 & 0.00e+00 & 0.00e+00 & 1.39e-08 & 6.75e-08 & 16.620 \\ 
             ros\_500 & 4.22e-11 & 0.00e+00 & 0.00e+00 & 0.00e+00 & 8.85e-08 & 3.87e-07 & 2.039 \\ 
              rose15 & 6.74e-08 & 0.00e+00 & 0.00e+00 & 0.00e+00 & 1.34e-02 & 2.68e-02 & Failed \\ 
         sensor\_1000 & 9.41e-10 & 0.00e+00 & 0.00e+00 & 0.00e+00 & 3.75e-07 & 5.58e-07 & 76.390 \\ 
          sensor\_500 & 7.91e-10 & 0.00e+00 & 0.00e+00 & 0.00e+00 & 5.06e-07 & 8.71e-07 & 19.840 \\ 
              shmup4 & 5.17e-06 & 0.00e+00 & 0.00e+00 & 0.00e+00 & 3.63e-07 & 4.81e-07 & 93.490 \\ 
              shmup5 & 2.73e-05 & 0.00e+00 & 0.00e+00 & 0.00e+00 & 3.64e-07 & 6.92e-07 & 668.800 \\ 
    spar060-020-1\_LS & 1.63e-08 & 0.00e+00 & 0.00e+00 & 0.00e+00 & 9.06e-08 & 9.73e-08 & 109.300 \\ 
           swissroll & 1.53e-03 & 0.00e+00 & 0.00e+00 & 0.00e+00 & -1.39e-02 & 7.32e-01 & Failed \\ 
              taha1a & 1.04e-01 & 0.00e+00 & 0.00e+00 & 0.00e+00 & 2.03e-02 & 1.56e+00 & Failed \\ 
              taha1b & 1.64e-11 & 0.00e+00 & 0.00e+00 & 0.00e+00 & 1.97e-07 & 4.15e-04 & 139.700 \\ 
              taha1c & 9.63e-01 & 0.00e+00 & 0.00e+00 & 0.00e+00 & 3.52e-01 & 1.81e-02 & Failed \\ 
            theta102 & 2.50e-10 & 0.00e+00 & 0.00e+00 & 0.00e+00 & 7.40e-09 & 1.13e-08 & 705.800 \\ 
             theta12 & 3.00e-10 & 0.00e+00 & 0.00e+00 & 0.00e+00 & 6.94e-09 & 8.56e-09 & 182.600 \\ 
       tiger\_texture & 6.12e-08 & 0.00e+00 & 0.00e+00 & 0.00e+00 & 1.57e-06 & 1.77e-03 & 245.300 \\ 
               trto4 & 1.52e-04 & 0.00e+00 & 0.00e+00 & 0.00e+00 & 6.81e-08 & 4.16e-07 & 6.302 \\ 
               trto5 & 6.89e-04 & 0.00e+00 & 0.00e+00 & 0.00e+00 & 3.61e-06 & 5.38e-06 & 75.200 \\ 
              vibra4 & 6.05e-05 & 0.00e+00 & 0.00e+00 & 0.00e+00 & 2.25e-07 & 3.59e-07 & 24.370 \\ 
              vibra5 & 2.83e-04 & 0.00e+00 & 0.00e+00 & 0.00e+00 & 1.19e-06 & 1.54e-06 & 314.000 \\ 
              yalsdp & 1.65e-10 & 0.00e+00 & 0.00e+00 & 0.00e+00 & 4.84e-07 & 4.88e-07 & 93.430 \\  
      \bottomrule
    \end{tabular}
  }  
\end{center}  
\end{table}}}

{\small{\begin{table}[h]
  \begin{center}
    \caption{COPT: Mittelmann's Benchmark Test Part 1}
  \label{hans-copt-1}
  \resizebox{\textwidth}{!}{%
  \begin{tabular}{cccccccc}
      \toprule
      Instance & Error 1 & Error 2 & Error 3 & Error 4 & Error 5 & Error 6 &
      Time\\
\midrule
            1dc.1024 & 2.04e-08 & 0.00e+00 & 0.00e+00 & 0.00e+00 & -5.73e-10 & 9.19e-08 & 196.160 \\ 
            1et.1024 & 1.50e-09 & 0.00e+00 & 0.00e+00 & 0.00e+00 & -3.92e-11 & 1.04e-08 & 37.670 \\ 
            1tc.1024 & 1.87e-08 & 0.00e+00 & 0.00e+00 & 0.00e+00 & -5.02e-10 & 1.38e-07 & 31.410 \\ 
            1zc.1024 & 2.98e-08 & 0.00e+00 & 0.00e+00 & 0.00e+00 & -1.17e-09 & 1.64e-07 & 63.790 \\ 
AlH\_1-Sigma+\_STO-6GN14r20g1T2\_5 & 7.34e-10 & 0.00e+00 & 0.00e+00 & 0.00e+00 & 2.50e-09 & 4.56e-09 & 1835.230 \\ 
            Bex2\_1\_5 & 2.51e-07 & 0.00e+00 & 0.00e+00 & 0.00e+00 & -4.07e-10 & 2.54e-06 & 5.630 \\ 
BH2\_2A1\_STO-6GN7r14g1T2 & 8.39e-10 & 0.00e+00 & 0.00e+00 & 0.00e+00 & 2.54e-09 & 7.04e-09 & 30.500 \\ 
               biggs & 7.32e-07 & 0.00e+00 & 6.03e-09 & 0.00e+00 & -3.51e-09 & 2.13e-06 & 1.760 \\ 
           broyden25 & 3.81e-08 & 0.00e+00 & 0.00e+00 & 0.00e+00 & -7.77e-10 & 1.95e-06 & 12.460 \\ 
         Bst\_jcbpaf2 & 8.55e-09 & 0.00e+00 & 0.00e+00 & 0.00e+00 & 1.56e-10 & 1.13e-07 & 7.270 \\ 
               buck4 & 4.17e-06 & 0.00e+00 & 0.00e+00 & 0.00e+00 & -2.94e-09 & 7.70e-06 & 41.940 \\ 
               buck5 & 2.16e-05 & 0.00e+00 & 0.00e+00 & 0.00e+00 & -6.07e-09 & 4.48e-05 & 834.240 \\ 
             butcher & 1.28e-03 & 0.00e+00 & 0.00e+00 & 0.00e+00 & -1.10e-05 & 3.39e-01 & Failed \\ 
          cancer\_100 & 4.10e-05 & 0.00e+00 & 0.00e+00 & 0.00e+00 & -8.40e-09 & 2.66e-06 & 38.400 \\ 
CH2\_1A1\_STO-6GN8r14g1T2 & 9.90e-10 & 0.00e+00 & 4.64e-10 & 0.00e+00 & 3.68e-09 & 7.75e-09 & 27.450 \\ 
         checker\_1.5 & 6.02e-10 & 0.00e+00 & 0.00e+00 & 0.00e+00 & 2.48e-09 & 2.57e-09 & 868.370 \\ 
            chs\_5000 & 1.09e-10 & 0.00e+00 & 0.00e+00 & 0.00e+00 & 7.40e-11 & 2.59e-09 & 39.060 \\ 
             cnhil10 & 2.20e-06 & 0.00e+00 & 0.00e+00 & 0.00e+00 & 4.10e-07 & 2.09e-05 & 3.430 \\ 
             cphil12 & 1.58e-11 & 0.00e+00 & 0.00e+00 & 0.00e+00 & 3.90e-12 & 6.03e-10 & 14.960 \\ 
       diamond\_patch & 7.08e-07 & 0.00e+00 & 1.15e-06 & 0.00e+00 & -1.65e-07 & 7.68e-04 & 2939.110 \\ 
e\_moment\_quadknap\_17\_100\_0.5\_2\_2 & 2.78e-07 & 0.00e+00 & 0.00e+00 & 0.00e+00 & -2.51e-09 & 1.57e-05 & 7.450 \\ 
e\_moment\_stable\_17\_0.5\_2\_2 & 7.00e-08 & 0.00e+00 & 0.00e+00 & 0.00e+00 & 2.64e-09 & 3.21e-06 & 5.670 \\ 
                foot & 7.32e-06 & 0.00e+00 & 0.00e+00 & 0.00e+00 & -7.37e-09 & 7.76e-06 & 245.410 \\ 
              G40\_mb & 8.27e-07 & 0.00e+00 & 0.00e+00 & 0.00e+00 & 3.45e-09 & 4.97e-06 & 98.970 \\ 
               G40mc & 5.06e-11 & 0.00e+00 & 0.00e+00 & 0.00e+00 & 1.72e-10 & 1.41e-08 & 80.450 \\ 
              G48\_mb & 7.27e-11 & 0.00e+00 & 0.00e+00 & 0.00e+00 & 1.53e-09 & 5.07e-08 & 183.950 \\ 
               G48mc & 1.12e-12 & 0.00e+00 & 0.00e+00 & 0.00e+00 & -4.22e-10 & 6.32e-10 & 118.790 \\ 
               G55mc & 1.48e-10 & 0.00e+00 & 0.00e+00 & 0.00e+00 & -3.67e-09 & 4.28e-09 & 1026.100 \\ 
               G59mc & 4.67e-10 & 0.00e+00 & 0.00e+00 & 0.00e+00 & -3.86e-11 & 7.64e-09 & 1131.240 \\ 
              G60\_mb & 2.09e-07 & 0.00e+00 & 0.00e+00 & 0.00e+00 & 9.02e-09 & 4.96e-07 & 5188.110 \\ 
               G60mc & 2.09e-07 & 0.00e+00 & 0.00e+00 & 0.00e+00 & 9.02e-09 & 4.96e-07 & 5189.290 \\ 
H3O+\_1-A1\_STO-6GN10r16g1T2\_5 & 7.27e-10 & 0.00e+00 & 0.00e+00 & 0.00e+00 & 3.49e-09 & 5.11e-09 & 93.900 \\ 
                hand & 4.68e-07 & 0.00e+00 & 0.00e+00 & 0.00e+00 & -2.50e-09 & 2.34e-07 & 49.390 \\ 
             ice\_2.0 & 3.03e-10 & 0.00e+00 & 0.00e+00 & 0.00e+00 & 3.90e-08 & 4.01e-08 & 8561.740 \\ 
            inc\_1200 & 1.10e-07 & 0.00e+00 & 0.00e+00 & 0.00e+00 & -1.25e-09 & 4.89e-06 & 99.020 \\ 
             mater-5 & 9.00e-07 & 0.00e+00 & 0.00e+00 & 0.00e+00 & -9.24e-10 & 1.60e-05 & 2.840 \\ 
             mater-6 & 1.08e-06 & 0.00e+00 & 0.00e+00 & 0.00e+00 & -3.08e-10 & 1.64e-05 & 7.100 \\ 
           neosfbr25 & 1.50e-09 & 0.00e+00 & 0.00e+00 & 0.00e+00 & 8.65e-10 & 6.25e-09 & 85.810 \\ 
         neosfbr30e8 & 9.48e-09 & 0.00e+00 & 0.00e+00 & 0.00e+00 & 4.62e-09 & 4.07e-08 & 438.710 \\ 
                neu1 & 2.58e-09 & 0.00e+00 & 2.91e-09 & 0.00e+00 & -8.26e-09 & 2.31e-06 & 3.610 \\ 
               neu1g & 7.58e-07 & 0.00e+00 & 0.00e+00 & 0.00e+00 & -2.01e-09 & 9.33e-07 & 3.000 \\ 
               neu2c & 2.81e-07 & 0.00e+00 & 1.41e-07 & 0.00e+00 & -7.33e-10 & 6.68e-07 & 7.560 \\    
               \bottomrule
    \end{tabular}
  }  
\end{center}
\end{table}}}

{\small{\begin{table}[h]
  \begin{center}
  \caption{COPT: Mittelmann's Benchmark Test Part 2 \label{hans-copt-2}}
  \resizebox{\textwidth}{!}{%
  \begin{tabular}{cccccccc}
      \toprule
      Instance & Error 1 & Error 2 & Error 3 & Error 4 & Error 5 & Error 6 &
      Time\\
      \midrule
                      neu2 & 7.66e-09 & 0.00e+00 & 1.19e-08 & 0.00e+00 & -4.11e-10 & 8.68e-08 & 2.980 \\ 
               neu2g & 1.65e-07 & 0.00e+00 & 3.05e-10 & 0.00e+00 & -1.30e-09 & 3.31e-07 & 4.340 \\ 
                neu3 & 2.15e-08 & 0.00e+00 & 0.00e+00 & 0.00e+00 & 8.48e-09 & 7.05e-07 & 19.410 \\ 
               neu3g & 1.44e-08 & 0.00e+00 & 0.00e+00 & 0.00e+00 & 9.38e-09 & 2.84e-07 & 24.360 \\ 
  NH2-.1A1.STO6G.r14 & 1.77e-09 & 0.00e+00 & 0.00e+00 & 0.00e+00 & 2.55e-09 & 7.60e-09 & 25.890 \\ 
NH3\_1-A1\_STO-6GN10r16g1T2\_5 & 5.22e-10 & 0.00e+00 & 0.00e+00 & 0.00e+00 & 2.98e-09 & 4.26e-09 & 94.540 \\ 
 NH4+.1A1.STO6GN.r18 & 1.48e-08 & 0.00e+00 & 0.00e+00 & 0.00e+00 & 2.09e-09 & 9.87e-09 & 591.090 \\ 
            nonc\_500 & 1.60e-09 & 0.00e+00 & 1.65e-10 & 0.00e+00 & 3.20e-10 & 2.60e-09 & 0.340 \\ 
         p\_auss2\_3.0 & 1.47e-08 & 0.00e+00 & 2.26e-14 & 0.00e+00 & 3.58e-09 & 3.58e-09 & 12868.900 \\ 
               rabmo & 1.36e-07 & 0.00e+00 & 0.00e+00 & 0.00e+00 & -1.63e-09 & 1.74e-05 & 5.150 \\ 
             reimer5 & 1.54e-06 & 0.00e+00 & 0.00e+00 & 0.00e+00 & -2.10e-09 & 3.84e-04 & 14.440 \\ 
    rendl1\_2000\_1e-6 & 9.69e-07 & 0.00e+00 & 0.00e+00 & 0.00e+00 & 3.37e-09 & 8.81e-07 & 111.430 \\ 
            ros\_2000 & 5.85e-10 & 0.00e+00 & 5.52e-11 & 0.00e+00 & 3.74e-10 & 2.78e-09 & 1.460 \\ 
             ros\_500 & 1.80e-09 & 0.00e+00 & 5.12e-10 & 0.00e+00 & 8.15e-10 & 1.25e-08 & 0.280 \\ 
              rose15 & 1.02e-08 & 0.00e+00 & 5.60e-10 & 0.00e+00 & -9.33e-09 & 9.29e-07 & 3.180 \\ 
         sensor\_1000 & 9.62e-09 & 0.00e+00 & 1.27e-08 & 0.00e+00 & 1.40e-10 & 5.41e-08 & 38.960 \\ 
          sensor\_500 & 5.85e-08 & 0.00e+00 & 0.00e+00 & 0.00e+00 & 4.41e-10 & 1.91e-07 & 9.570 \\ 
              shmup4 & 2.70e-07 & 0.00e+00 & 0.00e+00 & 0.00e+00 & -4.62e-09 & 2.36e-05 & 469.190 \\ 
              shmup5 & 1.78e-06 & 0.00e+00 & 0.00e+00 & 0.00e+00 & -6.45e-09 & 6.97e-05 & 7118.160 \\ 
    spar060-020-1\_LS & 5.90e-07 & 0.00e+00 & 0.00e+00 & 0.00e+00 & 4.96e-10 & 2.11e-06 & 11.910 \\ 
           swissroll & 2.95e-03 & 0.00e+00 & 0.00e+00 & 0.00e+00 & -2.68e-05 & 2.37e-02 & Failed \\ 
              taha1a & 2.46e-09 & 0.00e+00 & 0.00e+00 & 0.00e+00 & 6.98e-10 & 5.31e-08 & 5.030 \\ 
              taha1b & 5.56e-09 & 0.00e+00 & 6.74e-10 & 0.00e+00 & 1.28e-09 & 6.38e-08 & 21.220 \\ 
              taha1c & 7.48e-09 & 0.00e+00 & 0.00e+00 & 0.00e+00 & 1.90e-09 & 2.37e-07 & 38.820 \\ 
            theta102 & 4.25e-09 & 0.00e+00 & 0.00e+00 & 0.00e+00 & -4.15e-10 & 8.05e-09 & 484.170 \\ 
             theta12 & 2.87e-09 & 0.00e+00 & 0.00e+00 & 0.00e+00 & -1.25e-10 & 6.44e-09 & 60.580 \\ 
       tiger\_texture & 7.48e-07 & 0.00e+00 & 1.08e-06 & 0.00e+00 & -6.88e-08 & 1.14e-04 & 117.210 \\ 
               trto4 & 5.40e-05 & 0.00e+00 & 0.00e+00 & 0.00e+00 & -1.90e-08 & 3.83e-05 & 27.160 \\ 
               trto5 & 1.95e-04 & 0.00e+00 & 0.00e+00 & 0.00e+00 & -2.34e-08 & 1.52e-04 & 391.830 \\ 
              vibra4 & 1.14e-05 & 0.00e+00 & 0.00e+00 & 0.00e+00 & -6.30e-08 & 1.94e-04 & 64.660 \\ 
              vibra5 & 5.93e-05 & 0.00e+00 & 0.00e+00 & 0.00e+00 & -3.75e-07 & 3.78e-03 & 841.620 \\ 
              yalsdp & 8.63e-10 & 0.00e+00 & 8.86e-11 & 0.00e+00 & -4.99e-10 & 1.45e-09 & 9.460 \\ 
      \bottomrule
    \end{tabular}
  }  
\end{center}  
\end{table}}}

\end{document}